\documentclass[12pt]{article}
\usepackage{mheck}

\institution{SISSA}{\ Theoretical Particle Physics Group, SISSA, Trieste, 34100 Italy}
\institution{HARVARD}{\   Jefferson Physical Laboratory, Harvard University, Cambridge, MA 02138, USA}

\title{$Y$ systems, $Q$ systems, and 4D $\cn=2$ supersymmetric QFT}
\authors{Sergio Cecotti \worksat{\SISSA} \footnote{e-mail: {\tt cecotti@sissa.it}}
 and Michele Del Zotto \worksat{\HARVARD} \footnote{e-mail: {\tt eledelz@gmail.com}} }

\abstract{We review the connection of $Y$-- and $Q$--systems with the BPS spectra of $4D$
$\cn=2$ supersymmetric QFTs. For each finite BPS chamber of a $\cn=2$ model which is UV superconformal, one gets a periodic
$Y$--system, while for each finite BPS chamber of an asymptotically--free $\cn=2$ QFT one gets a $Q$--system
\textit{i.e.}\! a rational recursion all whose solutions satisfy a linear recursion with constant coefficients (depending on the
initial conditions). For instance, the classical $ADE$ $Y$--systems of Zamolodchikov correspond to the $ADE$ Argyres--Douglas $\cn=2$ SCFTs,
while the usual $ADE$ $Q$-systems to pure $\cn=2$ SYM. After having motivated the correspondence both from the QFT and the TBA sides, 
and having introduced the basic tricks of the trade,
we exploit the connection to construct \emph{and solve} new $Y$-- and $Q$--systems.

In particular, we present the new $Y$--systems associated to the $E_6,E_7,E_8$ Minahan--Nemeshanski SCFTs and to the $D_2(G)$ SCFTs.
We also present new $Q$--system corresponding to SYM coupled to specific matter systems such that the YM $\beta$--function remains negative.
\bigskip

\begin{flushright}
\emph{Dedicated to the memory of H.S.M. Coxeter}
\end{flushright}
}

\begin{document}

\maketitle

\tableofcontents


\section{Introduction and Overview}

$Y$--systems\footnote{ For a recent review see \cite{kuniba2}.} \cite{Zamo,tateo,kuniba}
 and $Q$--systems \cite{kirilov,kedem,difrancesco} are important tools in the theory of integrable systems which, 
rather surprisingly, turn out to unify structures fundamental for many areas of theoretical physics\footnote{ An important development, 
not discussed in the present paper, is the role of $Y$--systems in the holographic computation of Wilson loop expectation values at strong 
coupling, see ref.\cite{maldacena}.}
 and pure mathematics. 
Particularly interesting and useful is the physical interpretation of general $Y$-- and $Q$--systems in the context of QFT's 
(in diverse dimensions) having extended supersymmetry  \cite{CNV,Arnold1,Arnold2}. This interpretation puts these systems and their `magical'
properties in a new and broader perspective. The relation with supersymmetric QFT has paved the way to the construction and the 
explicit solution of a large number of new $Y$-- and $Q$--systems whose existence was not suspected before. 
In this paper we review such developments, providing also additional examples of $Y$-- and $Q$--systems, not published before,
 as well as outlining the method to construct and solve more general ones. \medskip

Our first basic claim \cite{CNV} is that there is a (family of) periodic $Y$--systems
 for each \emph{finite} BPS chamber of a $4D$ $\cn=2$ supersymmetric QFT which arises as the mass--deformation
 of a non--trivial $\cn=2$ SCFT. A BPS chamber is finite if it contains only finitely many hypermultiplets and no higher spin BPS particle.   
The original $ADE$ $Y$--systems defined by Zamolodchikov \cite{Zamo}  correspond,  
in this $\cn=2$ perspective, to the minimal\footnote{ \emph{Minimal} means the chamber with the smaller
 number of particles. For $ADE$ AD this corresponds to one hypermultiplet per simple root of 
the Lie algebra $\mathfrak{g}\in ADE$.} BPS chambers of the Argyres--Douglas (AD) models, which 
 are also classified by the simply--laced Lie algebras $ADE$ \cite{eguchi,CV11}. Likewise, the $Y$--systems 
associated to pairs $(G,G^\prime)$ of Dynkin graphs \cite{tateo} correspond to the canonical finite BPS chambers
 of the $(G,G^\prime)$ $\cn=2$ QFT's constructed\footnote{ In \cite{CNV} only the case with $G, G^\prime$ simply--laced was discussed in detail.
In general $G,G^\prime$ may be non--simply laced Dynkin graphs or even tadpole graphs \cite{tateo}. These $Y$--systems may be obtained
by folding the simply--laced ones along a group of `outer automorphisms', as in the examples of Appendix C of \cite{CNV}.
In this review we shall consider only the generalization of the simply--laced $Y$--systems, limiting ourselves to stressing 
that other non--simply--laced $Y$--systems may be constructed out of them by the folding procedure. We must admit that the 
corresponding issue is not 
adequately understood for the non--simply--laced $Q$--systems, despite the progresses of \cite{nonsimply}.} in \cite{CNV}. 
Other periodic $Y$--systems
 were constructed in \cite{Arnold1,Arnold2}, and several new examples
will be provided in the present paper.

In all these instances, the fact that the $Y$--system is periodic is a reflection of the fact 
that the associated $\cn=2$ QFT has a good UV scaling limit. Let us explain how this happens from 
the point of view of QFT. One starts in the UV with the given $\cn=2$ SCFT which has an exact $U(1)_R$ 
symmetry part of the $\cn=2$ superconformal algebra. Adding the mass deformation, we break explicitly (and softly) 
the $U(1)_R$ symmetry. In the BPS sector this breaking manifests itself as follows: from the \textsc{susy} algebra
\begin{equation}
\{Q^A_\alpha,Q^B_\beta\}= \epsilon_{\alpha\beta}\,\epsilon^{AB}\, Z
\end{equation} 
we see that the central charge $Z$ would transform under the $U(1)_R$ rotation $Q_\alpha^A\to e^{i\phi}\, Q_\alpha^A$ 
as $Z\to e^{2i\phi}\,Z$; but, as its name implies, $Z$ is \emph{central} in the Lie algebra 
of the continuous (unbroken) symmetries of the QFT \cite{hls},
 and \emph{no} continuous symmetry may rotate its phase. Since in a mass--deformed theory $Z$ is not the zero operator,
 we get a contradiction unless $U(1)_R$ is softly broken by the mass. However, a \emph{discrete} subgroup of $U(1)_R$ may still
 survive the deformation. A subgroup $\mathbb{Z}_\ell\subset U(1)_R$ survives precisely when the deforming operator 
$\mathscr{O}_\text{mass}$ commutes with
$\exp(2\pi i R/\ell)$, where $R$ is the SCFT $U(1)_R$ charge normalized so that $R(Q^A_\alpha)=1/2$. The operator
$\mathscr{L}\equiv\exp(2\pi i R/\ell)$ is then a symmetry of the massive QFT
which is called the $1/\ell$--fractional quantum monodromy \cite{CNV}.
The operator $\mathscr{M}=\exp(2\pi i R)$ is always a symmetry of the QFT (at least in the IR);
it is called the (full) quantum monodromy \cite{CNV}. The $1/2$--fractional monodromy $\mathscr{K}$  
is also well--defined for all $\cn=2$ models \cite{CNV}, since there always exists a symmetry inverting
 the sign of the \textsc{susy} central charge, $Z\leftrightarrow-Z$, namely PCT.
Whenever a $1/\ell$--fractional monodromy $\mathscr{L}$ exists, we have $\mathscr{M}=\mathscr{L}^\ell$, 
and the largest allowed $\ell$ must be even by PCT. The eigenvalues of the adjoint action of
the $\ell$--th fractional quantum monodromy are equal to
$\exp(2\pi i q_i/\ell)$, where $q_i$ are the $U(1)_R$ charges of the operators in the undeformed SCFT. 
Suppose --- as it happens in all known $\cn=2$ theories --- that the $q_i$'s are rational numbers of the
form $n/r$ with $n\in\mathbb{N}$ and a \textit{fixed} denominator $r$. Then one has
\begin{equation}
\mathscr{L}^{r\ell}=\mathscr{M}^r=1.
\end{equation} 
If $r$ is the smallest positive integer for which this equation is true, 
we say that the quantum monodromy has order $r$. 
As we review in \S.3.3, whenever we know the Seiberg--Witten geometry of the $\cn=2$ QFT to
 get the two integers $\ell$ and $r$ is a simple exercise in geometry \cite{CNV, Arnold1,CDZG}.

The action of $\mathscr{L}$ may be computed in a different way. Each (massive) BPS state carries a
 phase $e^{i\theta}$, namely the phase of its (non--zero) central charge $Z$. 
This phase specifies the subalgebra of the \textsc{susy} algebra which leaves invariant the BPS 
(short) supermultiplet. In a $\cn=2$ theory there are also half--BPS 
line--operators $\mathscr{O}_\gamma(\zeta)$, labelled by elements $\gamma$ of the lattice $\Gamma$ of conserved charges,
 which are invariant under the sub--supersymmetry
 of twistor parameter $\zeta\in\mathbb{P}^1$ \cite{GMN:2008,GMN:2009,GMN:2010,GMN:2011}.
The theory is quantized in the Euclidean space $\R^3\times S^1_R$, and the line operators $\mathscr{O}_\gamma(\zeta)$ are wrapped on the circle
$S^1_R$ of length $R$.
The vacuum expectation values of these operators $X_\gamma(\zeta)=\langle\mathscr{O}_\gamma(\zeta)\rangle$
are functions on a $4m$--dimensional hyperK\"ahler variety $\mathscr{T}$ which are holomorphic 
in complex structure $\zeta$ \cite{GMN:2008}. Here $m$ is the rank of the lattice $\Gamma$ of conserved charges. The $X_\gamma(\zeta)$'s
 jump when their phase $\zeta/|\zeta|$ aligns with the phase $e^{i\theta}$ of a BPS state \cite{GMN:2008}. 
The total variation of $X_\gamma(\zeta)$ as $\zeta\to e^{2\pi i/\ell}\zeta$ is given by the composition of 
all the jumps due to BPS states with phases in the wedge $\alpha\leq \theta <\theta+2\pi/\ell$  times a
 kinematic operator $L$, with $L^\ell=1$, which realizes the $\mathbb{Z}_\ell$--symmetry on the BPS spectrum 
(\textit{i.e.}\! it maps the BPS particles of phase $\theta$ into the BPS particles of phase $\theta+2\pi/\ell$) \cite{CNV}. 
The Kontsevich--Soibelman wall--crossing formula 
\cite{KS1}\!\!\cite{GMN:2008,wallcrossingtopstrings} states that each jump is a rational symplectic 
transformation of the $\{X_\gamma(\zeta)\}_{\gamma\in\Gamma}$. Therefore, if in the wedge 
$\alpha\leq \theta <\alpha+2\pi/\ell$ there are just \emph{finitely many} BPS particles, the action of $\mathscr{L}$ on
the $X_\gamma(\zeta)$'s is given by a finite composition of Kontsevich--Soibelman rational symplectomorphisms, times the 
linear map associated to the kinematic operation $L$; in conclusion,
$\mathscr{L}$ acts on the $X_\gamma(\zeta)$ as a rational map $\mathsf{L}\colon \mathbb{P}^m\to \mathbb{P}^m$.
 The $Y$--system associated to the $\ell$--th fractional monodromy $\mathscr{L}$ is then the rational recursion
\begin{equation}\label{LLLL1}
X_\gamma(\zeta)_n=\mathsf{L}\!\big(X_{\gamma^\prime}(\zeta)_{n-1}\big)_\gamma \
 \ \Bigg(\equiv X_\gamma(e^{2\pi i n/\ell}\zeta)\Bigg)\qquad n\in\Z.
\end{equation}   
The SCFT statement $\mathscr{L}^{r\ell}=1$ then translates into the statement that the $Y$--system is periodic in $n$ of period $r\ell$
\begin{equation}
X_\gamma(\zeta)_{n+r\ell}=X_\gamma(\zeta)_n\quad\forall\, n\in\Z,
\end{equation}
where $r$ and $\ell$ are the integers predicted by the Seiberg--Witten geometry. 
Up to diagram--folding, all known periodic $Y$--systems arise in this way for some finite BPS chamber of some $\cn=2$ theory, 
and in all instances we have full agreement between the Seiberg--Witten geometry and the properties of the 
solutions to the rational recursion relation \cite{CNV,Arnold1}.

From the previous discussion, we see that the operators $\mathscr{M}$, $\mathscr{K}$ (and $\mathscr{L}$ when defined)
are properties of the SCFT independent (up to conjugacy) of the particular mass--deformation. Hence two $Y$--systems defined by two different (finite)
chambers hasing the same $\mathbb{Z}_\ell$ symmetry are equivalent up to conjugation in the Cremona group. This fact is equivalent to
 the Kontsevich--Soibelman formula.

The idea of the quantum monodromy, and its relation with the jumps at the BPS phases $e^{i\theta_i}$,
 originates from the analysis of the $(2,2)$ $2d$ theories in \cite{CV92,CV91}. In the $2d$ case the quantum 
monodromy $H$ is (typically) a finite matrix whose eigenvalues are $\exp(2\pi i(q_i-\hat c/2))$, the $q_i$ 
being the $U(1)_R$ charges of the chiral primary fields of the UV SCFT. The situation becomes a bit subtler \cite{CV92}
when the $(2,2)$ theory is not UV conformal, but just asymptotically--free, \textit{i.e.}\! when we have logarithmic violations 
of the UV scaling. In that case $H$ is not diagonalizable, but has non--trivial Jordan blocks.
This is natural in view of the equality of the $U(1)_R$ charge and the scaling dimension for chiral primary operators.
 In the chiral sector the scaling operator is $\exp[\log\mu (\log H)/2\pi i]$ and the nilpotent part of $(\log H)/2\pi i$  is responsible
for producing the powers of $\log\mu$ which violate the UV scaling.
Equivalently, the nilpotent part of $(\log H)/2\pi i$ produces a mixing along the RG flow
 of the various chiral operators. The nilpotency 
(\textit{i.e.}\! triangularity) of 
this operation corresponds to the fact that a given operator $\mathscr{O}$ may mix only with operators $\mathscr{O}^\prime$ of lower 
UV dimension, $d(\mathscr{O}^\prime) < d(\mathscr{O})$ which have the same quantum numbers. 
Thus, acting on a chiral operator $\mathscr{O}$ of finite UV dimension, the minimal unipotent power of the monodromy 
$H^r$ produces $\mathscr{O}+\cdots$ where the ellipsis stands for a linear combination of
chiral operators of  lesser UV dimension. In a non--degenerate theory there are only finitely many such operators 
of dimension $\leq d(\mathscr{O})$; let $k(\mathscr{O})$ be the number of operators $\mathscr{O}^\prime$ 
with the same conserved quantum numbers of $\mathscr{O}$
with UV dimension $d(\mathscr{O}^\prime)\leq d(\mathscr{O})$. 
Then all the iterated (adjoint) actions of $H^r$,
\begin{equation}\label{arg0}
(H^r)^n\cdot\mathscr{O}\equiv \mathscr{O}_n\quad\text{with }\ n\in\mathbb{N},
\end{equation}
produce
$\mathscr{O}$ modulo  operators which belong to a
vector space of dimension (at most) $k-1=k(\mathscr{O})-1$.
Therefore, for all $n\in\Z$ the set of $(k+1)$ operators
\begin{equation}\label{arg1}
\big\{\,\mathscr{O}_{n+k},\ \ 
\mathscr{O}_{n+k-1},\ \
\mathscr{O}_{n+k-2},\ \
\cdots,\ \
\mathscr{O}_{n+1},\ \
\mathscr{O}_n\,\big\},   
\end{equation}
  satisfy a linear relation
  \begin{equation}
  \mathscr{O}_{n+k}= 
c_1(n)\,\mathscr{O}_{n+k-1}+
c_2(n)\,\mathscr{O}_{n+k-2}+
\cdots +c_{k-1}(n)\,\mathscr{O}_{n+1}+ c_k(n)\,\mathscr{O}_n=0.   
  \end{equation} 
Act with $H^r$ on both sides of this equality. We get
 \begin{equation}
  \mathscr{O}_{n+k+1}= 
c_1(n)\,\mathscr{O}_{n+k}+
c_2(n)\,\mathscr{O}_{n+k-1}+
\cdots +c_{k-1}(n)\,\mathscr{O}_{n+2}+ c_k(n)\,\mathscr{O}_{n+1}=0,   
  \end{equation}
that is, $c_i(n+1)=c_i(n)$ for all $i$ and the $c_i$'s are independent of $n$. 
We conclude that the family of operators $\mathscr{O}_n$ satisfy a finite--length linear recursion relation with \emph{constant} coefficients
\begin{equation}\label{argn}
\mathscr{O}_{n+k+1}= 
c_1\,\mathscr{O}_{n+k}+
c_2\,\mathscr{O}_{n+k-1}+
\cdots +c_{k-1}\,\mathscr{O}_{n+2}+ c_k\,\mathscr{O}_{n+1}=0.
\end{equation}
We must also have $c_k=\pm 1$; this is a consequence of CPT which implies that the operators
 $H$ and $H^{-1}$ are conjugate. Of course, in $2d$ all these statements (with an explicit formula for $k$ and 
the $c_i$'s) may be obtained by far more direct considerations \cite{CV92}. But the argument in eqns.\eqref{arg0}--\eqref{argn} 
makes sense, at least formally, also for the four--dimensional quantum monodromy $\mathscr{M}$ and its fractional powers
 $\mathscr{L}$ (if defined).
The argument will apply to $4D$ susy--protected \emph{local} operators. It is plausible that it extends also to the protected
half--BPS line operators
$\{\mathscr{O}_\gamma(\zeta)\}_{\gamma\in\Gamma}$ wrapped on the $S^1_R$ circle of the $\R^3\times S_R^1$ geometry, since
 they look local operators
from the $3d$ perspective.  

We are thus lead to expect that in an \emph{asymptotically--free} $4D$ $\cn=2$ QFT, which has a $\mathbb{Z}_\ell$--symmetric
 \emph{finite} BPS chamber,
the 
$n$--th successive iterations
of the
action of the rational transformation $\mathsf{L}$, corresponding to the $\ell$--th fractional monodromy $\mathscr{L}$ 
eqn.\eqref{LLLL1},
\begin{equation}
X_\gamma(\zeta)_n\equiv \big\langle\,\mathscr{L}^n\cdot\mathscr{O}_\gamma(\zeta)\,\big\rangle =
 \mathsf{L}^n\big(X_{\gamma^\prime}(\zeta)\big)_{\gamma},\qquad n\in\Z 
\end{equation}
 will satisfy a finite linear recursion of the form ($s=r\ell$)
\begin{equation}\label{receeers}
X_\gamma(\zeta)_{n+ks}=\sum_{j=1}^{k-1} c_j(\gamma,\zeta)\, X_{\gamma}(\zeta)_{n+js}\pm X_\gamma(\zeta)_n,\quad\text{for all }n\in\Z,
\end{equation} 
where the coefficients $c_j(\gamma,\zeta)$ are independent of $n$ (but may depend on everything else). This peculiar expectation
is confirmed by several explicit examples; a number of highly non--trivial checks are presented in section 5 of this review.

Note that, while the integers $r$, $\ell$, and $s=k\ell$ are universal, depending only on the $\cn=2$ model,
the degree of the linear recursion $k=k(\mathscr{O}_\gamma)$ depends on the particular operator considered. 
The minimal $k(\mathscr{O}_\gamma)$ behaves roughly\footnote{ \emph{Roughly} means the following: if 
$k(\mathscr{O}_\gamma)$, $k(\mathscr{O}_{\gamma^\prime})$ are the minimal degrees of the recursion relations for $X_\gamma$,
$X_{\gamma^\prime}$, then $X_{\gamma+\gamma^\prime}$ satisfies a recursion relation of degree
 $k(\mathscr{O}_\gamma)\cdot k(\mathscr{O}_{\gamma^\prime})$
which \emph{generically} is minimal but not always so. Then 
$k(\mathscr{O}_{\gamma+\gamma^\prime})\leq k(\mathscr{O}_\gamma)\cdot k(\mathscr{O}_{\gamma^\prime})$. For instance,
(generically) $k(\mathscr{O}_{2\gamma})=k(\mathscr{O}_\gamma)^2-k(\mathscr{O}_\gamma)+1$.} multiplicatively
\begin{equation}
 k(\mathscr{O}_{\gamma+\gamma^\prime})= k(\mathscr{O}_\gamma)\cdot k(\mathscr{O}_{\gamma^\prime}).
\end{equation}
The numbers $r$ and $\ell$ are easily extracted from the Seiberg--Witten curve; to get the minimal $k(\mathscr{O}_\gamma)$ requires more work.

We define a $Q$--system to be a non--linear rational map $X_{\gamma,n}\xrightarrow{\ \mathsf{L}\ } X_{\gamma, n+1}$ 
such that its $n$--fold iterations satisfy --- for all variables $\{X_\gamma(\zeta)\}_{\gamma\in\Gamma}$ and \emph{all} initial conditions ---  
a finite--length linear recursion of the form \eqref{receeers}. 
The classical examples of $Q$--systems are the $ADE$ ones in \cite{difrancesco}.\smallskip

Then our second basic claim is that there is a (family of) $Q$--systems associated to each finite
 BPS chamber of an \emph{asymptotically--free} $4D$ $\cn=2$ theory. In particular, the $ADE$ 
$Q$--systems of ref.\cite{difrancesco} correspond to the minimal (strongly coupled) chamber of 
pure $\cn=2$ SYM with gauge group equal to the $ADE$ group in question. 
In this paper we present several new examples of $Q$--systems, and prove that their solutions satisfy linear recursion
 relations of the expected form \eqref{receeers}.

\medskip

The rest of the paper is organized as follows:
in section 2 we review the connection of $\cn=2$ QFT and $Y$--/$Q$--systems from the point of view of the 
TBA integral equations. In section 3 we review the quiver approach to the $\cn=2$ theories and their relations with the cluster algebras.
We also review the results of \cite{CNV} we need. In section 4 the $Y$--systems are discussed from the QFT perspective. As an illustration of the 
power of the \textsc{susy} viewpoint, we present the new $Y$--systems associated to some  non--trivial $\cn=2$ SCFT
namely the Minahan--Nemeschansky theories \cite{MN1,MN2} and the $D_2(G)$ model \cite{CDZG}.
In section 5 we consider the $Q$--systems and present new examples. In appendix A the $Y$--systems associated to 
$\cn=2$ models obtained by geometric engenering on a Arnol'd singularity \cite{Arnold1,Arnold2} are described.

\section{TBA, $\cn=2$ supersymmetry, $Y$-- and $Q$--systems}\label{TBAsection}

The notion of $Y$--system was introduced by Zamolodchikov in his analysis of the thermodynamical Bethe ansatz (TBA) for reflectionless
$ADE$  scattering theories \cite{Zamo}. There is one such theory for each simply--laced Lie algebra $\mathfrak{g}$; it contains
 $r=\mathrm{rank}\, \mathfrak{g}$ particles species with masses $m_a$, $a=1,\dots, r$. The pseudoenergies $\varepsilon_a(\beta)$ of the 
various species are determined,  as a function of the rapidy $\beta$, by the TBA integral equations \cite{Zamo}
\begin{equation}\label{zamozamo}
 \varepsilon_a(\beta)= R\, m_a\, \phi(\beta)+\frac{h}{2\pi}\sum_b l_{ab}\int_{-\infty}^{+\infty} 
\frac{d\beta^\prime}{2\,\cosh\!\big(h (\beta^\prime-\beta)/2\big)}\, \log\!\Big(1+\exp\!\big[\varepsilon_b(\beta^\prime)\big]\Big),
\end{equation}
 where $R$ is the spatial size of the system, $h$ is the Coxeter number of $\mathfrak{g}$, and $l_{ab}$ is expressed in terms 
of its Cartan matrix $C_{ab}$ as $l_{ab}=-C_{ab}+2\,\delta_{ab}$. $\phi(\beta)$ is some known universal function, and the masses satisfy
$l_{ab} m_b= 2 m_a \cos(\pi/h)$. The identity \cite{kuniba2}
\begin{equation}
 \frac{1}{4\,\cosh\!\big(\frac{\pi}{2} (u-v+i-i\epsilon)\big)}+\frac{1}{4\,\cosh\!\big(\frac{\pi}{2} (u-v-i+i\epsilon)\big)}=\delta(u-v)
\end{equation}
then implies that $Y_a(\beta)\equiv \exp[\varepsilon_a(\beta)]$ satisfy the equations
\begin{equation}
 Y_a(\beta+i\pi/h)\, Y_a(\beta-i\pi/h)= \prod_b\big(1+Y_b(\beta)\big)^{l_{ab}},
\end{equation}
so that the solutions to the TBA equations give particular solutions of the recursion relations
\begin{equation}\label{ADEysy}
 Y_a(n+1)\, Y_a(n-1)= \prod_b \big(1+Y_b(n)\big)^{l_{ab}},\qquad n\in\Z,
\end{equation}
which define the associated $Y$--system. Zamolodchikov conjectured, and Frenkel--Szenes \cite{frenkel} and, independently,
Gliozzi--Tateo \cite{gliozzi} proved that the solution to this recursion relation is periodic for \textit{all}
initial conditions, the period being (a divisor of) $2(h+2)$
\begin{equation}
Y_a(n+2h+4)=Y_a(n) \qquad \text{for all }n\in\Z. 
\end{equation}
Since the work of Zamolodchikov, many other periodic 
$Y$--systems have being constructed. For instance we have the $Y$--systems specified by a pair $(\mathfrak{g},\mathfrak{g}^\prime)$ of $ADE$ 
Lie algebras  \cite{tateo}
\begin{equation}\label{pairofG}
Y_{a,a^\prime n+1}\, Y_{a,a^\prime, n-1} =\frac{\prod_{b\in S} \big(1+Y_{b,a^\prime, n}\big)^{l_{ab}}}{\prod_{b^\prime\in S^\prime} 
\big(1+Y_{a,b^\prime, n}^{-1}\big)^{l^\prime_{a^\prime b^\prime}}},
\end{equation}
where $S$ (resp.\! $S^\prime$) is the set of simple roots of $\mathfrak{g}$ (resp.\! $\mathfrak{g}^\prime$) $a,b\in S$, $a^\prime, b^\prime 
\in S^\prime$, and $l_{ab}=2\delta_{ab}-C_{ab}$, $l^\prime_{a^\prime b^\prime}=2\delta_{a^\prime b^\prime}-C^\prime_{a^\prime b^\prime}$ 
in terms of the Cartan matrices $C, C^\prime$ of $\mathfrak{g}, \mathfrak{g}^\prime$.
The solutions $Y_{a,a^\prime,n}$ are periodic in $n$ of period a divisor of $2(h+h^\prime)$ \cite{kellerper}. 
For a recent review see \cite{kuniba2}.

On the other hand, Gaiotto, Moore and Neitzke, in their work \cite{GMN:2008} to understand the Kontsevich--Soibelman wall--crossing formula
\cite{KS1} from the viewpoint of the hyperK\"ahler geometry of the effective $3d$ sigma--model, found that the
holomorphic Darboux coordinates $X_\gamma(\zeta)$ satisfy, as a function of the twistor coordinate $\zeta\in\mathbb{P}^1$,
the TBA--like system of integral equations \cite{GMN:2008,Gaiottoopers}
\begin{equation}
 \log X_\gamma(\zeta)= \frac{R}{\zeta} Z_\gamma+i\theta_\gamma + R \zeta\, \overline{Z}_\gamma+\sum_{\gamma^\prime}
\Omega(\gamma^\prime) \frac{\langle\gamma^\prime,\gamma\rangle}{4\pi i} 
\int_{\ell{\gamma^\prime}}\frac{d\zeta^\prime}{\zeta^\prime} \frac{\zeta^\prime+\zeta}{\zeta^\prime-\zeta}\; \log\!\Big(1-\sigma(\gamma^\prime) 
X_{\gamma^\prime}(\zeta^\prime)\Big)
\end{equation}
where the sum is over the stable BPS states of charge $\gamma^\prime\in \Gamma$, $\Omega(\gamma^\prime)$ is the net number (index) of
BPS states with charge $\gamma^\prime$, $\langle\cdot,\cdot \rangle \colon \Gamma\otimes \Gamma\to \Z$ is the Dirac
 electro--magnetic antisymmetric 
pairing, and 
$\sigma(\gamma^\prime)$ assumes the values $\pm 1$
(see \S.3 below for details on the QFT notations). These equations resemble very much Zamolodchikov equations \eqref{zamozamo}. 
The relation between the two becomes more transparent if, in some chamber, we have a \emph{bi--partite}
BPS spectrum, that is,
 we can split the set of the charge
vectors $\{\gamma\}$ of stable BPS particles in two disjoint subsets 
\begin{equation}
\{\gamma\}=\{\gamma\}_\text{odd}\cup \{\gamma\}_\text{even} 
\end{equation}
so that
the odd (resp.\! even) BPS particles are mutually local (that is, the Dirac pairing $\langle\cdot,\cdot\rangle$ vanishes
 when restricted to either subset of definite parity). Then setting
\begin{equation}\label{gaiogaio}
 Y_\gamma(\zeta)= \begin{cases}
                   -\sigma(\gamma)\,X_\gamma(\zeta) & \gamma\ \text{odd}\\
-\sigma(\gamma)\,X_\gamma(-i\zeta) & \gamma\ \text{even},
                  \end{cases}
\end{equation}
working on the locus $\theta_\gamma=0$. where we have the additional symmetry \cite{Gaiottoopers}
\begin{equation}\label{extrasym}
Y_{\gamma}(-\zeta)=Y_{-\gamma}(\zeta)\equiv Y_{\gamma}(\zeta)^{-1},
\end{equation}
 and combining together the contribution from the BPS state of charge
$\gamma^\prime$ with the one from its PCT conjugate of charge $-\gamma$, we get \cite{Gaiottoopers}
\begin{equation}
 \log Y_\gamma(\beta)=R\,\phi_\gamma(\beta)+ \frac{h}{2\pi} \sum_{\gamma^\prime>0} A_{\gamma,\gamma^\prime} \int\limits_{-\infty}^{+\infty} 
\frac{d\beta^\prime}{2\,\cosh\!\big(h (\beta^\prime-\beta)/2\big)}\;\log\!\Big(1+
Y_{\gamma^\prime}(\beta^\prime)\Big)
\end{equation}
 where $\zeta=e^{h\beta/2}$. Eqn.\eqref{gaiogaio} is formally identical to \eqref{zamozamo} provided:
 \textit{(i)} 
the integral matrices $l_{ab}$ and $A_{\gamma,\gamma^\prime}$ are 
equal, and \textit{(ii)} the functions $\phi_\gamma$ form an eigenvector of $A_{\gamma^\prime,\gamma}$ with the eigenvalue $2 \cos(\pi/h)$.

In $4d$ $\cn=2$ theory, the matrix $A_{\gamma,\gamma^\prime}$ depends on the BPS spectrum, hence on both the particular QFT model and
the BPS chamber. The simplest $\cn=2$ theories are the Argyres--Douglas (AD) models \cite{AD}
 which are classified by the simply--laced Lie algebras $ADE$
\cite{eguchi,CV11}.
The charge lattice of the model associated with the algebra $\mathfrak{g}$ is identified with the root lattice of $\mathfrak{g}$,
$\Gamma=\bigoplus_a \Z \alpha_a$,
and the Dirac pairing is $\langle \alpha_a,\alpha_b\rangle= (-1)^a l_{ab}$ (up to mutation, see \S.3). 
In its \emph{minimal} BPS chamber (the one with less stable BPS particles) the type $\mathfrak{g}$ AD theory has
 precisely one BPS hypermultiplet per simple root $\alpha_a$ 
(plus their PCT conjugates with charge $-\alpha_a$) \cite{CNV,ACCERV1}; in particular, the spectrum is bi--partite. 
Then $A_{\gamma,\gamma^\prime}$ reduces to the $l_{ab}$ matrix, and condition \textit{(i)} is satisfied. Condition \textit{(ii)} 
then selects the usual symmetric locus in the parameter space (the one studied in, say, \cite{shapere}).

In particular, in the AD models the Darboux coordinates $Y_a(\zeta)$ satisfy the $ADE$ $Y$--system \eqref{ADEysy}. What is the physical
motivation for this? The shift $n\to n+1$ in $Y_{a,n}$ is equivalent to $\beta\to \beta+i\pi/h$ in rapidity, or 
$\zeta\to e^{i\pi/2}\zeta$, which is a $U(1)_R$ rotation by $\pi/2$. The half--monodromy $\mathscr{K}$ acts on the central
charge as $Z\leftrightarrow -Z$ equivalent to $\zeta\leftrightarrow -\zeta$, or $\beta\to \beta+2 i\pi/h$. 
From eqns.\eqref{extrasym}\eqref{gaiogaio} we see that the kinematical operator $L$ associated to the $1/2$--fractional monodromy
$\mathscr{K}$ is just the inversion $I\colon Y_\gamma\leftrightarrow Y_{\gamma}^{-1}$ \cite{CNV}, while the \textsc{rhs}
of \eqref{ADEsysy} is the Kontsevich--Soibelman jump at the phase $\theta=\theta(\alpha_a)+i\pi/2$ where sit 
all $\alpha_b$--charge BPS particles 
which are not mutually local with respect to the $\alpha_a$--charge one. 
Therefore, in the AD case the action of the half--monodromy $\mathscr{K}$ is simply
\begin{equation}
 Y_{a,n} \longmapsto \mathscr{K}\cdot Y_{a,n} \equiv Y_{a,n+2}.
\end{equation}
The periodicity result just says that the 
quantum monodromy $\mathscr{M}$ has finite order, and in particular
$\mathscr{M}^{h+2}=1$, which is in agreement with the Seiberg--Witten geometry of the AD models \cite{CNV}. 
In particular, it is well known that the $U(1)_R$ charges have the form 
$n/(h+2)$  \cite{argyres}.

However, not all consistent $4d$ $\cn=2$ theories have a non--trivial UV SCFT fixed point.
The other possibility, for a consistent QFT, is to be just asymptotically--free (AF). 
In the UV the theory approaches its limiting theory with logarithmic
violation of scaling and $\mathrm{M}$ cannot be expected to have finite order.
As argued in the previous section, we expect finite--linear recursions to replace
periodicity (that is, we expect $Q$--systems instead of $Y$--systems).
Let us see how this happens by comparing the two simplest (non--trivial) examples of
the two possibilities: UV SCFT and AF. These models are given by, respectively,
type $A_2$ AD and pure $\cn=2$ SYM with gauge group $G=SU(2)$ whose integral
TBA equations have the same form as the ones for the AD models except that
the Cartan matrix of a finite--dimensional Lie algebra is replaced by the one of the affine Kac--Moody algebra
$A_1^{(1)}$ \cite{GMN:2008,Gaiottoopers,CV11}.
In particular, the integral equations for the second model are obatined from those of the first one
\eqref{gaiogaio} by the simple replacement $A_{\gamma,\gamma^\prime}\to 2\,A_{\gamma,\gamma^\prime}$. Using
the existence (in both cases) of a quarter--monodromy, we may rewrite the equations in the form \cite{Gaiottoopers}
\begin{align}\label{inttt1}
 &\log Y(\beta)= R\, \phi(\beta)+ \frac{1}{2\pi}\int_{-\infty}^{\infty} \frac{d\beta^{\prime}}{\cosh(\beta^\prime-\beta)}\; 
\log\!\Big(1+Y(\beta^\prime)\Big) &&\text{for }A_2\ \text{AD}\\
 &\log Y(\beta)= R\, \widetilde{\phi}(\beta)+ \frac{1}{\pi}\int_{-\infty}^{\infty} \frac{d\beta^{\prime}}{\cosh(\beta^\prime-\beta)}\; 
\log\!\Big(1+Y(\beta^\prime)\Big) &&\text{for }SU(2)\ \text{SYM}.
\label{inttt2}\end{align}
The UV limit is just $R\to 0$. If the theory has a good UV scaling limit, we expect that
$\log Y(\beta)$ has a smooth $R\to 0$ limit. Moreover, in the UV SCFT the $U(1)_R$ symmetry is exact and unbroken,
which means that v.e.v.\! $Y(\beta)$ must be independent of the phase of $\zeta$, that is, independent of $\mathrm{Im}\,\beta$.
By analyticity of the solution, we get the implication
\begin{equation}\label{secimplication}
 \text{UV SCFT}\ \Longrightarrow\ \lim_{R\to 0} Y(\beta) =Y\ \text{a constant independent of }\beta.
\end{equation}
Knowing that the solution $Y$ must be a constant, it is elementary to solve the $R\to 0$
limit of equation \eqref{inttt1}. Performing the integral we get $2\log Y= \log(1+Y)$, that is,
the algebraic equation
\begin{equation}\label{yyyverl}
 Y^2=Y+1
\end{equation}
which was interpreted in terms of RCFT characters and Verlinde algebras in \cite{CNV}. The splitting field of 
 \eqref{yyyverl} is the maximal totally real subfield of $\mathbb{Q}(e^{2\pi i/5})$, confirming the fact that the
$A_2$ $Y$--system is periodic of period $3+2=5$.

On the other hand, a constant solution to \eqref{inttt2} would have to satisfy
\begin{equation}
 Y=Y+1,
\end{equation}
which has no solutions. Hence, inverting the implication in \eqref{secimplication}, we conclude that the corresponding
$\cn=2$ QFT has to be AF, which is of course correct, since the $\beta$--function coefficient of pure $SU(2)$ $\cn=2$ SYM is $-4\neq 0$.
Indeed, the would--be $Y$ system of the $SU(2)$ SYM is
\begin{equation}\label{SU2Ysys}
 Y_{n+1}\,Y_{n-1}=(1+Y_n)^2.
\end{equation}
Writing $Y_n=X_n^2$ this reduces to the recursion relation for the Chebyshev polynomials (the characters of $SU(2)$), 
which is the basic example of a $Q$--system
\cite{difrancesco}
\begin{equation}\label{SU2frieze}
 X_{n+1}\,X_{n-1}=1+X_n^2\quad\Longleftrightarrow\quad \left|\begin{matrix} X_{n+1} & X_n\\
                                                              X_n & X_{n-1}
                                                             \end{matrix}
\right|=1,
\end{equation}
which implies
\begin{equation}
 0= \left|\begin{matrix} X_{n+2} & X_{n+1}\\
 X_{n+1} & X_{n}
\end{matrix}
\right|-\left|\begin{matrix} X_{n+1} & X_n\\
 X_n & X_{n-1}
\end{matrix}\right|
=\left|\begin{matrix} X_{n+2}+X_n & X_{n+1}\\
 X_{n+1}+X_{n-1} & X_{n}
\end{matrix}\right|
\end{equation}
which is equivalent to a $3$--terms linear recursion relation (with constant coefficients) of the form
\begin{equation}
 X_{n+2}= c\,X_{n+1}-X_n,\qquad \forall\, n\in\Z,
\end{equation}
where the constant $c$ depends on the initial conditions of the recursion. The physical variables $Y_n$ then satisfy the $4$--term
linear recursion
\begin{equation}
 Y_{n+3}= (c^2-1)Y_{n+2}-(c^2-1)Y_{n+1}+Y_n,
\end{equation}
as expected from the general arguments of the previous section. Those arguments suggest that this is the general pattern for all
asymptotically--free $\cn=2$ QFT. In this paper we shall corroborate these QFT predictions by checking the existence of a linear recursion
for several new $Q$--systems arising from AF $\cn=2$ QFT's.

One may wonder what happens in the third case, that is, in presence of UV Landau poles (\textit{i.e.}\! positive $\beta$--function).
In this case the theory is not UV complete, and hence not a consistent QFT on its own grounds. It may still be formally studied as a low--energy
 effective theory of some UV completion. The simplest example is the would--be $\cn=2$ theory associated to the hyperbolic Kac--Moody algebra
of Cartan matrix
\begin{equation}
 C=\begin{pmatrix}2 & - 3\\ - 3 & 2\end{pmatrix}
\end{equation}
(corresponding to the $3$--Kronecker quiver). As all formal $\cn=2$ models based on hyperbolic KM Lie algebras,
this theory is not UV complete \cite{CNV,cattoy}, 
but it has UV completions in the form of Minahan--Nemeshanski QFT's. In this case the only modification in the TBA equation would be
to put a factor $3$ in front of the  integral in eqn.\eqref{inttt1}. A constant solution $Y$ would then satisfy the algebraic equation
\begin{equation}
 Y^2=(1+Y)^3,
\end{equation}
 which is not consistent with any periodicity of the associated would--be $Y$--system
since its Galois group is non--Abelian (its discriminant $-23$ is not a square in
$\mathbb{Q}$). The modified integral equation
 is also not consistent with any finite--degree linear recursion of the form \eqref{receeers}. Indeed, a necessary
condition for a non--periodic sequence $Y_n$ to satisfy a finite--degree linear recursion with constant coefficients, the last one being $\pm1$,
is the existence of the following limit
\begin{equation}\label{pppqwe}
 \lim_{n\to \infty} \frac{Y_{n+1}}{Y_n}=\varrho >1,
\end{equation}
while the would--be $Q$--system gives
\begin{equation}
 \frac{Y_{n+1}}{Y_n}\cdot \frac{Y_{n-1}}{Y_n}= \frac{1}{Y_n^2}\big(1+Y_n\big)^3.
\end{equation}
Eqn.\eqref{pppqwe} would give $1$ as the $n\to\infty$ limit of the \textsc{lhs},
 while the \textsc{rhs} will go like $C \varrho^n$ for large $n$, giving a contradiction. This is hardly a surprise: in section 1 we argued for
a linear relation using the properties of the RG flow in the extreme UV of the QFT, and this flow 
ought to be pathological in a UV \emph{non}--complete theory.

\section{BPS spectral problem and cluster algebras}
\subsection{A lightning review of the BPS quiver approach}
\noindent{\bf Quiver representations and stability.} A quiver is a quadruple $Q \equiv (Q_0,Q_1,s,t)$ where $Q_0$ and $Q_1$ are two discrete sets ($Q_0$ is the set of nodes, $Q_1$ is the set of arrows) and $s,t: Q_1 \to Q_0$ are two maps that associate to an arrow its starting (resp. ending) node. A path of the quiver is a concatenation of arrows, composed as were functions. The set of all paths generates as a $\C$ vector space the path algebra of $\C Q$, the product of two path being their concatenation or zero (when concatenation is impossible). A complex representation of a quiver is a functor $X\colon \C Q \to \text{Vect}_{\C}$, where $\text{Vect}_{\C}$ is the category of complex vector spaces. $X$ associates to each node $i\in Q_0$ a vector space $X_i$ and to each arrow $\a \in Q_1$ a linear map $X_\a \colon X_{s(\a)} \to X_{t(\a)}$. The set of all complex quiver representations is itself an abelian category, $\text{rep}(Q)$. A morphism in this category $F \colon X \to Y$ is a collection of linear maps $(F_i)_{i\in Q_0}$ such that $F_{t(\a)} X_{\a} = Y_{\a} F_{s(\a)}$ for all $\a \in Q_1$. A morphism $F$ is injective if all linear maps $F_i$ are injective. As usual, subrepresentations and injective morphism are in bijection. Let $n$ be the number of nodes of $Q$. To each representation $X$ is associated a vector in $\mathbb{Z}_+^{n}$, the dimension vector of $X$, $\dim X \equiv (\dim_{\C} X_1,\dots,\dim_{\C} X_n)$. Let us denote by $\mathfrak{h}$ the complex upper half plane. A stability condition for $\text{rep}(Q)$ is a linear map $\zeta \colon \mathbb{Z}^n \to \C$ that associates to each dimension vector an element of $\mathfrak{h}$. By abuse of notation we are going to denote $\zeta(X)$ the quantity $\zeta(\dim X)$. A representation $X$ of $Q$ is said to be stable iff
\be
 0 \leq \arg \zeta(Y) < \arg \zeta(X) < \pi \quad \forall \ 0\neq Y \subset X.
\ee
Whenever a quiver has cycles, $i.e.$ whenever $Q$ contains a set of arrows $(\a_1,\dots,\a_\ell)$ such that $s(\a_{i+1}) = t(\a_{i})$, $i$ mod $\ell$, one can define a superpotential $\cw$ for $Q$ as a formal linear combination of such cycles. The jacobian ideal $\partial \cw$ is the set of cyclic derivatives of $\cw$ with respect to all the arrows $\a \in Q_1$. Given a pair $(Q,\cw)$, a representation of the quiver $Q$ with superpotential $\cw$ is a representation $X$ of $Q$ such that $X(\partial \cw) = 0$. All the above definitions can be carried over replacing $\text{rep}(Q)$ with the abelian category of representations of $(Q,\cw)$.

\medskip

\noindent{\bf BPS quivers of 4D $\cn=2$ gauge theories.}\cite{cattoy,DFR1,DFR2,FiolMarino,Fiol,Denef,review,ACCERV2} Consider a 4D $\cn=2$ gauge theory with a simply--laced gauge group $G$ of rank $r$.  Let such system flow to its infrared Coulomb branch: the gauge group breaks to $U(1)^r$. The system may also have a global flavor symmetry group of rank $f$ that for generic values of the mass deformations is $U(1)^f$. The electric, magnetic and flavor charges of the system being quantized are valued into an integer symplectic lattice $\Gamma$ of rank $n \equiv 2r +f$. The antisymmetric pairing on $\Gamma$ is the Dirac pairing in between electric and magnetic charges $\langle \cdot\,,\cdot \rangle \colon \Gamma \times \Gamma \to \mathbb{Z}$. The central charge of the 4D $\cn=2$ superalgebra $Z\equiv \epsilon^{\a \, \b} \epsilon_{A\,B} \{Q^A_{\a},Q^B_{\beta}\}$ gives a linear map $Z\colon \Gamma \to \C$ that depends on all parameters of the theory (Coulomb branch moduli, mass deformations, UV couplings, ...)\cite{OW}. Any given state of charge $\gamma \in \Gamma$ has mass greater than or equal to $|Z(\gamma)|$ (BPS bond). States that saturate this bond are called BPS. A 4D $\cn=2$ model has a BPS quiver iff there exist a set of generators $\{e_i\}$ of $\Gamma$ such that all charges of BPS states $\gamma \in \Gamma$ satisfy
\be\label{quivprop}
\gamma \in \Gamma_+ \quad\text{ or }\quad \gamma \in -\Gamma_+,
\ee
where $\Gamma_+ \equiv \oplus \, \mathbb{Z}_+ \, e_i$ is the (convex) cone of particles. The BPS quiver is encoded in the antisymmetric matrix $B_{ij} \equiv \langle e_i\,, e_j \rangle$. Its nodes are in one to one correspondence with the generators $\{e_i\}$ of $\Gamma$. If $B_{ij}\geq0$ one draws $B_{ij}$ arrows from node $i$ to node $j$. The dynamics on the worldline of a BPS state of charge $\gamma = \sum \, m_i \,e_i  \in \Gamma_+$ is encoded in a $\cn=4$ SQM of quiver type.\cite{Denef} The underlying quiver of this SQM is generically much more complicated than the BPS quiver. It is however a gauge theory with gauge group $\prod U(m_i)$ and arrows corresponding to bifundamental hypermultiplets. In particular, it can have loops (\emph{i.e.} adjoint hypers)  or 2 cycles (\emph{e.g.} mass or mixing terms). If $Q$ has cycles, the SQM has a superpotential obtained as a linear combination of single trace operators built out of the set of bifundamental hypers. If one studies the Higgs branch of this SQM, however, all adjoint hypers (loops) gets vev's and all loops in the quiver decouple. In the deep infrared moreover one expects to be able to integrate out all fields entering quadratically in the superpotential. The reduced quiver with the effective superpotential $\cw$ describing this regime of the SQM is the BPS quiver. The effective F terms give the Jacobian ideal we discussed in the previous paragraph, while the D terms can be traded for the stability condition and complexification of gauge groups by standard GIT arguments. Thus, any solution of the effective F term equations for the SQM associated to a BPS particle of charge $\gamma \in \Gamma_+$ is a representation of the BPS quiver $(Q,\cw)$ with dimension vector $\dim X = \gamma$. The stability condition is obtained from $Z(\cdot)$ as follows: for a generic $Z(\cdot)$ by the BPS quiver property we can always choose a phase $\theta \in [0,2\pi )$ such that $Z(\Gamma_+)$ lie in a positive convex cone inside $e^{i\,\theta}\mathfrak{h}$. Given a representation $X$ of $(Q,\cw)$, we define its stability conditions as $\zeta(X) \equiv e^{-i\theta} Z(X) \in \mathfrak{h}$. The SQM Higgs branch moduli space for a BPS particle with charge $\gamma=\sum \, m_i \,e_i  \in \Gamma_+ $ is the K\"ahler variety
\be 
\cm_\gamma \equiv \left\{\begin{gathered} X\in \text{rep}(Q,\cw) \, \text{stable} \\ \text{ and such that } \dim X = \gamma \end{gathered} \right\} \Big/ \prod GL(m_i,\C)
\ee
The $SU(2)_R \times \text{Spin}(2)$ dof's of the Clifford vacuum of our BPS state of charge $\gamma$ are encoded in the Dolbeault cohomology of $\cm_\gamma$. In particular, if $d = \dim_{\C} \cm_{\gamma}$, the spin is $(0,\frac{1}{2})\otimes \frac{d}{2}$. For example a given charge corresponds to a (half) hypermultiplet if the corresponding moduli space is a point, \emph{e.g.} the representation is rigid. Of course, the splitting in between particles and antiparticles in eqn.\eqref{quivprop} is artificial: \textsc{pct} symmetry acts on the charge lattice $\Gamma$ as a $\mathbb{Z}_2$ involution $\gamma \mapsto - \gamma$. As we are going to describe below, the BPS quiver associated to a 4D $\cn=2$ model with quiver property is not unique. The same non--uniqueness is mirrored by the 1d Seiberg--like duality groupoid of the 1d $\cn=4$ SQM.\footnote{ Here for simplicity of the discussion we are neglecting all the crucial subtleties related with the 1d SQM superpotential. In particular, we are implicitly assuming that all BPS quivers are 2 acyclic: While this is true for all the examples that we are going to discuss in the text, we stress here that this is not necessarily the case.\label{Wsubtle}}

\medskip

\noindent{{\bf Mutations.}}\cite{ACCERV2,CFIKV,DWZ,DWZ2} The choice of the (convex) cone of particles $\Gamma_+$ in the charge lattice $\Gamma$ fixes the BPS quiver of the model. Correspondingly there exists a $\theta \in [0,2\pi)$ such that $Z(\Gamma_+) \subset e^{i \theta} \mathfrak{h}$. Consider tilting the half plane $e^{i \theta} \mathfrak{h}$ clockwise. At $\theta=\theta_*$, the leftmost $Z(e_i)$ crosses the image of the negative real axis and for $\theta > \theta_*$ exits the half plane $e^{i \theta} \mathfrak{h}$. Correspondingly $Z(- e_i)$ will enter from the right side crossing the positive real axis. Notice that since $\Gamma_+$ is convex, it is always a generator of $\Gamma$ that exits the half plane $e^{i \theta} \mathfrak{h}$ in this way. Let us proceed a little bit further in the tilting and stop at a phase $\theta^\prime$ before any other such crossings occurs. The intersection of $Z(\Gamma)$ with $e^{i \theta^\prime}\mathfrak{h}$ defines implicitly a new convex cone of particles $\Gamma^\prime_+$ and a new set of generators $\{e^\prime_i\}$ with the quiver property. The two sets are related by a change of basis, the right charge lattice mutation at $i$-th node of the quiver:
\be
\mu_i(e_k)\equiv \begin{cases} - e_i &\text{for } k=i\\ e_k + [B_{ik}]_+ \,e_i &\text{else}\\  \end{cases}
\ee
where $[x]_+\equiv \text{max}(x,0)$. The BPS quiver has to be changed according to the Dirac pairing of the new set of generators. The $B$ matrix undergoes an elementary quiver mutation at node $i\in Q_0$:
\be\label{quivmut}
\mu_i (B_{k\, j}) = \begin{cases} - B_{k j}&\text{if }k=i \text{ or } j = i\\
B_{k \, j} + \text{sign}(B_{k \, i})[ B_{k \, i} \, B_{i \, j}]_+ &\text{otherwise} \end{cases}
\ee
For a generic superpotential $\cw$ this mutation rule is obtained from 1d SQM Seiberg like dualities.\footnote{ With the same \emph{caveat} of footnote \ref{Wsubtle}.} Tilting anti clockwise the plane, one obtains the left mutation, differing from the right mutation by a twist
\be
\widetilde{\mu}_i(e_k) \equiv \begin{cases} - e_i &\text{for } k=i\\ e_k + [-B_{ik}]_+ \,e_i &\text{else}\\  \end{cases}
\ee
Notice that the induced elementary quiver mutation at $i\in Q_0$ is the same in the two cases
\be
\widetilde{\mu}_i(B) \equiv \mu_i(B).
\ee
In particular the square of a right mutation is not an involution, but a non trivial transformation of the charge lattice, known as a Seidel Thomas twist: $\mu_i \circ \mu_i = t_i \neq  \text{id}$. While the inverse of a right mutation is the corresponding left mutation $\mu_i \circ \widetilde{\mu_i} = \text{id} =\widetilde{\mu_i} \circ \mu_i$. Below, when we loosely speak about a mutation of the charge lattice, we always mean a right mutation.

\medskip

Since physics is \textsc{pct} symmetric, the different choices of cones of particles $\Gamma_+$ leading to different quiver descriptions of the same physical system are all equivalent. To a given 4D $\cn=2$ model with quiver property is associated the whole mutation class of its BPS quivers.

\subsection{Wall crossing and quantum cluster algebras}\label{WC}

\noindent{{\bf Wall crossing and the spectral problem.}} The central charge of the 4D $\cn=2$ superalgebra depends on all parameters of the system. As we vary them it changes and so does the corresponding stability condition. In the process it might happen that the set of stable BPS particles changes. More precisely this happens whenever the central charges $Z$ of mutually non local BPS particles align. Let us describe this in more detail. Pairs of central charges of non local BPS states align along real codimension one loci in the space of all possible central charges $\C^n = (\Gamma\otimes \C)^{\vee}$. These loci are called walls of the first kind as opposed to walls of the second kind where a quiver mutation occurs. Walls of the first kind divide the space of central charges in domains $\{\cd_a\}_a$. In the interior of each such domain, the theory is in a different phase, characterized by a given spectrum of stable BPS states. The set of charges of such stable states is a BPS chamber in the charge lattice $\mathscr{C}_a \subset \Gamma$. Going across a wall of marginal stability a phase transition occurs and the BPS spectrum jumps from a BPS chamber to another: this is a wall crossing transition. The BPS spectral problem consist of finding the allowed pairs $\{\cd_a,\mathscr{C}_a\}$. The problem is simplified by the existence of a wall crossing invariant, the quantum monodromy or Kontsevich Soibelman operator, that encodes the spectrum.\cite{KS1} In order to solve the BPS spectral problem it is enough to compute the invariant in  one BPS chamber and its corresponding domain of the first kind, and this determines implicitly all other such pairs via wall crossing identities.\footnote{ Here we are neglecting a crucial physical fact: the whole of $(\Gamma\otimes \C)^{\vee}$ is not physical. Let us denote by $\mathcal{P}$ the space of physical parameters of a given model. The map $\mathcal{P} \to (\Gamma\otimes \C)^{\vee}$ that associates to a given set of values of the parameters of the theory $\{ \l \}$ a central charge function $Z_{\{\l\}}$ defines a subvariety of $(\Gamma\otimes \C)^{\vee}$ that for a generic model has positive codimension.\label{quantumShot}}

\medskip

\noindent{\bf Quantum monodromy.}\cite{CNV} Since the lattice $\Gamma$ is symplectic, it is canonically associated to a quantum torus algebra with elements $Y_\gamma$ for $\gamma \in \Gamma$. Let $q$ be a formal variable (the quantization parameter). The product of $\bT_\Gamma(q)$ is defined by 
\be\label{qtor}
Y_\gamma \,Y_{\gamma'} \equiv q^{\langle \gamma, \gamma' \rangle} \, Y_{\gamma'}\,Y_\gamma
\ee
The additive structure of the charge lattice can be lifted to the quantum torus via normal ordering, namely
\be\label{norm}
Y_{\gamma + \gamma'} \equiv N[Y_\gamma Y_{\gamma'}] \equiv q^{-\langle \gamma, \gamma' \rangle/2} Y_\gamma \, Y_{\gamma'}
\ee
that is associative and commutative. The 4D $\cn=2$ monodromy we have mentioned in the previous section is the inner automorphism of the quantum torus algebra:\footnote{ Here and below $\text{Ad}^\prime(x) a \equiv x^{-1} \, a \, x$.}
\be
Y_\c \longrightarrow \mathscr{M}_q^{-1} \, Y_\c \,\mathscr{M}_q \equiv \text{Ad}^\prime(\mathscr{M}_q) \, Y_{\c}
\ee
Let us proceed, following \cite{CNV}, to the construction of $\mathscr{M}_q$ in terms of the spectral data associated to a domain of the first kind $\cd_a$. Let $\mathscr{C}_a$ be the BPS chamber associated to $\cd_a$. For each $\c \in \mathscr{C}_a$, let $j_{\c}$ denote the higher spin of the corresponding Clifford vacuum. To any BPS particle $(\c\, , j_\c)$ corresponds an element of $\bT_{\Gamma}(q)$, 
\be
\co(\c\, , j_\c\, ; q) \equiv \prod_{s = -j_\c}^{j_\c} \Psi(-q^s \,Y_\c;q)^{(-1)^{2s}}
\ee
where the function $\Psi(x;\, q)$ is the quantum dilogarithm.\footnote{ The definition the we use here differs slightly from all other definitions in the litterature. Cfr. footnote 7 of \cite{CNV}. In the meantime yet another definition of quantum dilog appeared, the $\bE(Y)$ of \cite{Keller:dilog} related to the one used here by $\Psi(Y;q)= \bE(-Y)^{-1}$.}
The quantum dilogarithm \cite{Faddeev} is uniquely characterized by the two properties
\be\label{defdilog}
\begin{cases}\text{functional eqn}: &\Psi(qx ; \, q) = (1-q^{1/2} x)^{-1} \Psi(x ; \,q)\\
\text{normalization}:&\Psi(0;q) = 1
\end{cases}
\ee
Notice that
\be
\Psi(q^{-1}x;q) = (1+q^{1/2}x) \Psi(x;q) \qquad \Psi(x;q^{-1}) = (\Psi(x;q))^{-1}
\ee
If $q=\text{exp}(2 \pi i \tau)$ with $\tau$ in the upper half plane, we can solve eqn.\eqref{defdilog} in terms of a convergent infinite product
\be\label{dilogz}
\Psi(x;\,q) \equiv \prod_{n=0}^\infty \left(1-q^{n+1/2} x\right).
\ee 
In what follows we assume that $q$ is not a root of unity. The reader interested in the quantum Frobenius property is warmly advised to consult \cite{CNV}. For each charge $\gamma \in \Gamma$, let us define its `time' as $t_\gamma \equiv \arg Z(\gamma) \in S^1$. 
The quantum monodromy operator is the time ordered product
\be\label{KSop}
\mathscr{M}_q\equiv T \prod_{\c\in\mathscr{C}_a} \co(\c,j_\c;q).
\ee
The KS wall--crossing formula is the statement that the conjugacy class of this operator does not depend on the particular $\cd_a$ nor on the particular $\mathscr{C}_a$ we use to compute it: $\mathscr{M}_q$ is a wall--crossing \emph{invariant}. 

\medskip

Notice that if a chamber $\mathscr{C}_a$ has a $\mathbb{Z}_m$ involution,\emph{i.e.} a linear transformation $\mathbf{V}: \mathscr{C}_a \to \mathscr{C}_a$ such that $\mathbf{V}^m = \text{id}$, by linearity of the central charges, the corresponding domain of the first kind inherits such a symmetry. $\mathscr{C}_a$ splits into $m$ (identical) subchambers $\mathscr{S}_a$: $\mathscr{C}_a=\coprod_{k=0}^{m-1} \mathbf{V}^k \mathscr{S}_a$. Lifting the action of $\mathbf{V}$ to an adjoint action on the quantum torus $\bT_\Gamma(q)$ yields the involution
\be
\mathscr{V} \, Y_\gamma \, \mathscr{V}^{-1} \equiv Y_{\mathbf{V} \, \gamma}.
\ee
In this case, the quantum monodromy $\mathscr{M}_q$ has a $1/m$ root
\bea\label{frak}
\mathscr{M}_q \equiv \mathscr{Y}_q^m
\eea
where
\be\label{fracdefn}
\mathscr{Y}_q \equiv \mathscr{V}^{-1} \circ \, T \prod_{\c\in\mathscr{S}_a} \co(\c\, , j_\c\, ; q)
\ee
is the $1/m$ fractional quantum monodromy operator\cite{CNV,Arnold1}. Notice that by wall crossing identities the splitting in eqn.\eqref{frak} is always true, even in other chambers where the original symmetry might be broken. Since physics is \textsc{pct} symmetric, the $1/2$ fractional  monodromy $\mathscr{K}_q$ is always well defined.

\medskip

If one can find a representation $\ch$ of $\bT_{\Gamma}(q)$, the statement that the conjugacy class of $\mathscr{M}_q$ is wall--crossing invariant implies the invariance of the trace $\text{Tr}_\ch(\mathscr{M}_q)$. A very nice representation of $\bT_{\Gamma}(q)$ was found in \cite{wallcrossingtopstrings} that allows to interpret $\text{Tr}_{\ch} \mathscr{M}_q$ as a topological partition function. Here we closely follow \cite{IV13}. Assume the model has a nontrivial UV superconformal point where its global $U(1)_R$ symmetry is restored, let $R$ be the corresponding charge. A Melvin cigar is a 3-manifold $MC_q$ defined as a quotient of $\C \times S^1$ with respect to the relation $(z,\theta) \sim (qz,\theta + 2 \pi)$. Now, let $H_R$ denote the generator of the Cartan of the $SU(2)_R$ symmetry of the 4D $\cn=2$ model. Consider the topologically twisted theory on the background $MC_q \times_R S^1$, where the $R$-twist is given by the identification of the $R-H_R$ charge with the holonomy around this second $S^1$. Notice that by doing this we are breaking 4 of the 8 supercharges of 4D $\cn=2$ \textsc{susy}. Let us define
\be
Z(t,q) \equiv \text{Tr}_{MC_q \times_R S^1} (-1)^F t^{R-H_R} \equiv \langle (-1)^F t^{R-H_R} \rangle_{MC_q \times_R S^1}
\ee
In this background one obtains a representation of $\bT_{\Gamma}(q)$ from the reduction of the BPS--line operators of the 4D $\cn=2$ theory down to the 1d theory on the $S^1$ $R$--circle.\footnote{ For a more detailed and intriguing description of this story we refer to the original paper \cite{CNV}} Moreover,
\be\label{blackmagic}
\text{Tr}_{\ch}\,[\mathscr{M}_q]^k = Z(t=e^{2\pi i k},q) = \langle (-1)^F \text{exp}(2 \pi k R) \rangle_{MC_q \times_R S^1}
\ee
If the SCFT has $R$--charges of the form $\mathbb{N}/\ell$, we \emph{predict} via \eqref{blackmagic}, that $\text{Ad}[\big(\mathscr{M}_q\big)^\ell] = \text{id}_{\bT_{\Gamma}(q)}$. This fact can be used to give a very non trivial check of all of these ideas, through the duality in between 4D $\cn=2$ SCFT and periodic TBA $Y$--systems discussed in section \S.\ref{TBAsection}.

\medskip

The construction of the wall crossing invariants is simplified for the models that have a finite chamber of hypermultiplets via the quantum cluster algebra associated to the pair $(Q,\cw)$.

\medskip

\noindent{\bf Quantum mutations.}\cite{BZ,qd-cluster,clqd2,qd-pentagon} Choose a set of generators $\{e_i\}$ with the BPS quiver property: to each generator corresponds a generator $Y_i \equiv Y_{e_i}$ of the quantum torus $\bT_\Gamma(q)$. In particular, by eqn.\eqref{qtor},
\be\label{genqtor}
Y_i \, Y_j \equiv q^{B_{ij}} Y_j \, Y_i,
\ee
and $Y_{-e_i} \equiv (Y_i)^{-1}$. As previously, while the expression in eqn.\eqref{qtor} is a mutation invariant, the one in eqn.\eqref{genqtor} is not. The mutations we have introduced in the previous section are lifted via the normal ordering \eqref{norm} to the generators of the quantum torus algebra
\be\label{wrzk}
\mu_i (Y_k) \equiv Y_{\mu_i(e_k)} \qquad \widetilde{\mu}_i (Y_k) \equiv Y_{\widetilde{\mu}_i(e_k)}
\ee
An elementary quantum mutation $\cq_i$ at the node $i\in Q_0$ is
\be\label{quantumut}
\cq_i \cdot Y_k \equiv \text{Ad}^\prime (\Psi(-Y_i;\,q)) \circ \mu_i (Y_k) \equiv \Psi(-Y_i;\,q)^{-1}\, \mu_i(Y_k) \,\Psi(-Y_i;\,q)
\ee
Or, more explicitly:
\be\label{explicitquantumut}
\cq_i \cdot Y_j = \begin{cases} \prod_{n = 0}^{B_{ij}-1} (1+q^{n+1/2} Y_i)^{-1} q^{- (B_{ij})^2/2} \,Y_j \, Y_i^{B_{ij}} &\text{if } B_{ij}\geq 1\\
 \prod_{n = 0}^{|B_{ij}|-1} (1+q^{-n-1/2} Y_i)\, Y_j &\text{if } B_{ij}\leq - 1 \\
 (Y_i)^{-1} & \text{if } i=j\\
 Y_j&\text{else}
\end{cases}
\ee
by eqns.\eqref{norm}, \eqref{dilogz} and \eqref{quantumut}. Let us remark that
\be
\Psi(-Y_i;\,q)^{-1}\, \mu_i(Y_k) \,\Psi(-Y_i;\,q) = \widetilde{\mu}_i \big(\Psi(-Y_i;\,q)\,Y_k\, \Psi(-Y_i;\,q)^{-1}\big)
\ee
by explicitly evaluating the RHS and comparison with eqn.\eqref{explicitquantumut}. Thus
\be
\cq_i \equiv \text{Ad}^\prime (\Psi(-Y_i;\,q)) \circ \mu_i  \equiv \widetilde{\mu}_i \circ \text{Ad}(\Psi(-Y_i;\,q)) 
\ee
is an involution of $\bT_\Gamma(q)$, namely: $\cq_i \circ \cq_i = \text{id}_{\bT}$.

\medskip

\noindent{\bf The mutation method.} The mutation method of \cite{ACCERV2} is a very convenient algorithm to compute $\bK(q)$. Let $\Lambda\equiv\{i_1,i_2,\dots,i_{p-1},i_p\}$ be an ordered sequence of nodes in $Q_0$. Let us denote $\mathbf{m}_\Lambda$ the sequence of  mutations
\be
\mathbf{m}_\Lambda \equiv \mu_{i_p} \circ \mu_{i_{p-1}}\circ \cdots \circ \mu_{i_2}\circ \mu_{i_1}.
\ee
A sequence of mutations is said to be admissible iff $i_k \neq i_{k+1}$ for all $k=1,...,p-1$.
Consider a 4D $\cn=2$ model with BPS quiver. Assume that it has a finite chamber of hypermultiplets $\mathscr{C}_\text{fin}$. Let  $\{e_i\} \subset \mathscr{C}_\text{fin}$ be a set of generators of the charge lattice satisfying the quiver property. As discussed above, this choice defines a cone of particles $\Gamma_+$ inside $\Gamma$, and an angle $\theta$ such that $Z(\Gamma_+) \subset e^{i\theta} \mathfrak{h}$. Consider rotating $e^{i\theta} \mathfrak{h}$ clockwise. Each time the image $Z(e_{i_k})$ of a generator exits from the rotating half plane, the basis of the charge lattice has to be changed accordingly, by the right mutation $\mu_{i_k}$ at the $i_k$-th node of $Q$. Let us continue to vary $\theta$ and to mutate the quiver accordingly. By \textsc{pct} symmetry, $e^{i(\theta \pm \pi)} \mathfrak{h}\cap Z(\Gamma) = - Z(\Gamma_+)$. Therefore we obtain back the original quiver $Q$: A finite BPS chamber consisting of $p$ hypermultiplets defines a finite ordered sequence of nodes $\Lambda \equiv \{i_1,\dots,i_p\}$ such that
\begin{itemize}
\item $i_k \neq i_{k+1}$ for all $k=1,...,p-1$ (admissible);
\item $\mathbf{m}_\Lambda(Q) = \pi(Q)$ where $\pi \in \mathfrak{S}_{n}$ is a permutation of the labels of the nodes  (automorphism of $\bT_\Gamma(q)$);
\item $\mathbf{m}_\Lambda(e_k) \equiv -e_{\pi(k)}$ (\textsc{pct} symmetry) ;
\end{itemize}
\noindent The corresponding $p$ charge vectors of particles in $\mathscr{C}_\text{fin}$ are
\be\label{thecharges}
\gamma_m = \mathbf{m}_{\{i_1,i_2,\dots,i_{m-1}\}}(e_{i_m})\in \Gamma^+ \qquad m=1,\dots,p
\ee
ordered in decreasing phase: $\arg Z(\gamma_m) > \arg Z(\gamma_{m+1})$.\footnote{ We choose conventions in such a way that BPS time flows clockwise in the definition of $\mathscr{M}$.} Notice that by $\mathbf{m}_{\Lambda}(Q) = Q$, the corresponding mutation sequence is lifted to an automorphism of the quantum torus algebra: this is precisely the \textsc{pct} transformation $-\text{id}_\Gamma$ twisted by the permutation $\pi$:
\be\label{pcts}
\mathbf{m}_{\Lambda}(Y_i) = Y_{-\pi(e_i)} \equiv \text{Ad}(\mathscr{V}_\textsc{pct}\, \mathscr{I}_\pi) \, Y_{i}
\ee
Obviously, $[\mathscr{I}_\pi,\mathscr{V}_\textsc{pct}] = 0$. Reversing the logic of the construction, each admissible ordered sequence of nodes $\Lambda$ satisfying
\be\label{thesoln}
\mathbf{m}_\Lambda(e_k) = - e_{\pi(k)}
\ee
gives a solution of the BPS spectral problem. As we are going to discuss below, the method can be refined if the finite chamber has more symmetries. Notice that if we `quantize' the above sequence of mutations we obtain
\bea
\cq_{\Lambda} &\equiv \cq_{i_{p}} \circ  \cq_{i_{p-1}} \circ \cdots \circ  \cq_{i_2} \circ  \cq_{i_1}\\
&= \text{Ad}^\prime(\Psi(-Y_{i_p};\,q)) \circ \mu_{i_p} \circ \text{Ad}^\prime(\Psi(-Y_{i_{p-1}};\, q)) \circ \mu_{i_{p-1}} \circ\cdots\\
&\qquad\qquad\cdots\circ \text{Ad}^\prime(\Psi(-Y_{i_2};\, q))\circ \mu_{i_2} \circ \text{Ad}^\prime(\Psi(-Y_{i_1};\, q))\circ \mu_{i_1}\\
& = \text{Ad}^\prime\Big(\Psi(-Y_{e_{i_1}};\,q)\Psi(-Y_{\mu_{i_1}(e_{i_2})};\,q)\cdots\\
&\qquad\qquad\cdots\Psi(-Y_{\mathbf{m}_{i_1,\dots,i_{p-3}}(e_{i_{p-2}})};\,q) \Psi(-Y_{\mathbf{m}_{i_1,\dots,i_{p-2}}(e_{i_{p-1}})};\,q)\Big)\circ \mathbf{m}_{\Lambda}\\
&= \text{Ad}^\prime(\mathscr{I}_{\pi^{-1}} \circ \mathscr{K}_q).
\eea
We have just showed that the quantization of the mutation sequence given by the mutation method computes the adjoint action of the half--monodromy on $\bT_\Gamma(q)$.

\medskip

The same method can be generalized easily to compute the fractional monodromy. Since in this case we have a $\bZ_m$ involution $\mathbf{V}\colon \mathscr{C} \to \mathscr{C}$, the image $Z(\mathscr{C})$ can be divided into $m$ distinct angular sectors (and compatibility with $\textsc{pct}$ entails that $m$ is even). If $\mathscr{C}$ is a finite BPS chamber is sufficient to rotate $e^{i\theta} \mathfrak{h}$ clockwise till $e^{i(\theta - 2\pi/m)} \mathfrak{h}$ to capture a mutation sequence that, once quantized, computes the adjoint action of the $1/m$ fractional monodromy on $\bT_\Gamma(q)$. In other words we obtain a sequence of nodes $\Xi=\{k_1,\dots,k_s\} \subset Q_0$ that satisfies
\begin{itemize}
\item $k_j \neq k_{j+1}$ for all $j=1,...,s-1$ (admissible);
\item $\mathbf{m}_\Xi(Q) =\sigma(Q)$ where $\sigma \in \mathfrak{S}_{n}$ ;
\item For compatibility with \textsc{pct} symmetry, $\exists \, k_\Xi \in \mathbb{Z}_{\geq 0}$ such that $\mathbf{m}_\Xi$ satisfies 
\be\label{thebettersoln}
\mathbf{m}_{\sigma^{k_{\Xi}}(\Xi)} \circ\cdots \circ \mathbf{m}_{\sigma^2(\Xi)} \circ\mathbf{m}_{\sigma(\Xi)} \circ \mathbf{m}_\Xi(e_i) = - e_{\pi(i)}
\ee
\end{itemize}
\noindent Again, reversing the logic, any admissible sequence of nodes $\Xi$ that maps the quiver into itself up to a permutation of the nodes $\sigma \in \mathfrak{S}_n$ such that eqn.\eqref{thebettersoln} holds corresponds to a quantum fractional monodromy of order $2 (k_{\Xi}+1)$. The same steps as above yields
\be
\cq_{\Xi} = \cq_{k_s} \circ \cq_{k_{s-1}} \circ \cdots \circ \cq_{k_1}= \text{Ad}^\prime(\mathscr{I}_{\sigma^{-1}} \circ \mathscr{Y}_q).
\ee
Here we have used the fact that since $\mathbf{m}_\Xi(Q)=Q$ up to a permutation, $\mathbf{m}_\Xi$ lifts to an automorphism of $\bT_\Gamma(q)$, that is precisely $\mathbf{V}$ twisted by $\sigma$:
\be
\mathbf{m}_\Xi(Y_{i}) =\text{Ad}(\mathscr{V}\mathscr{I}_\sigma) \, Y_i.
\ee
Of course, $[\mathscr{V},\mathscr{I}_\sigma]=0$.

\medskip

\noindent{\bf Factorized--sequences of mutations.}\label{sinksourcess}\cite{BGP,kellerper,Arnold1} A node of a quiver $i \in Q_0$ is called a \emph{source} (resp.\emph{sink}) if there is no arrow $\a \in Q_1$ such that $t(\a) = i$ (resp.$s(\a)=i$). Let us notice here that with our conventions a right (resp.left) mutation on a node $i$ that is a sink (resp.source) have the only effect of reversing the sign of the generator $e_i$ leaving all the other generators unchanged. A sequence of nodes $\Lambda=\{i_1,i_2,\cdots, i_k\}$ of a quiver $Q$ is called a \emph{sink} sequence (resp.\! a \emph{source} sequence) it the $i_s$ node is a sink (resp.\! a source) in the mutated quiver $\mu_{i_{s-1}}\mu_{i_{s-2}}\cdots \mu_{i_1}(Q)$ for all $1\leq s\leq k$. A sink (resp.\! source) sequence $\Lambda$ is called \emph{full} if contains each node of $Q$ exactly once. If $Q$ is acyclic, and $\Lambda$ is a full source sequence, the corresponding right mutation sequence acts on the generators of the charge lattice as the Coxeter element $\Phi_Q\colon \Gamma \to \Gamma$, while $\Lambda^{-1}$ is a full sink sequence and the corresponding right mutation sequence acts as the inversion. 

\bigskip

Given a subset $S$ of the set of nodes $Q_0$, we introduce the notation $Q|_S$ to denote the full subquiver of $Q$ over the nodes $S$. Consider the node set $Q_0$ as the disjoint union of a family of sets $\{q_\a\}_{\a\in A}$:
\begin{equation}
Q_0 = \coprod_{\a \in A} q_\a
\end{equation}
To each subset of nodes $q_\a$ we associate the full subquiver $Q|_{q_\a}$ of $Q$. Given a node $i \in Q_0$, we will denote $q_{\a(i)}$ the unique element in the family that contains node $i$.

\bigskip

Now, consider an admissible finite sequence of nodes $\Lambda$ such that $\mathbf{m}_\Lambda$ satisfies eqn.\eqref{thesoln}. $\Lambda$ is said to be \emph{source-factorized} of type $\{Q|_{q_\a}\}_{\a \in A}$ if
\begin{itemize}
\item[$i)$] For all $\ell = 1,2,...,m$, the $\ell$-th node in the sequence $i_\ell$ is a sink in 
\begin{equation}
\mu_{i_{\ell-1}} \circ \cdots \circ \mu_{i_1}(Q)\big|_{\{i_\ell\} \cup Q_0 \setminus q_{\a(i_\ell))}}
\end{equation}
\item[$ii)$] For all $\ell = 1,2,...,m$ the $\ell$-th node in the sequence $i_\ell$ is a source in
\begin{equation}
\mu_{i_{\ell-1}} \circ \cdots \circ \mu_{i_1}(Q)\big|_{q_{\a(i_\ell))}}
\end{equation}
\end{itemize}
In our conventions source factorized sequences of mutations are appropriate for right mutations. For the dual left mutations one shall invert sources with sinks in $i)$ and $ii)$ above.\footnote{ For saving time and print in view of the applications that we have in mind, here we give just a simplified version of the story: the interested reader is referred to the original paper \cite{Arnold1} for the whole beautiful story.}

\medskip

A source factorized sequence is in particular \emph{Coxeter-factorized} of type $(Q|_{q_\a})_\a$, provided all $Q|_{q_\a}$ are Dynkin $ADE$ quivers with alternating orientation. If this is the case, let us denote by $G_{\a}$ the alternating quiver $Q|_{q_\a}$. A sequence of right mutations that is Coxeter--factorized is automatically a solution of \eqref{thesoln}. In particular, if all the alternating $ADE$ subquivers of the family are ALL equal to a given $G$, one has a $1/m$ fractional monodromy of order $m=\text{lcm}(2,h(G))$.

\medskip

For a Coxeter--factorized source--sequence, by the right mutation rule combined with $i),ii)$, each element of the sequence acts as the simple Weyl reflection 
\begin{equation}
s_{i_\ell} \in \text{Weyl}(Q\big|_{q_{\a(i_\ell))}})
\end{equation}
on the charges on nodes $i \in q_{\a(i_\ell))}$, and as the identity operation on all other charges. By the standard properties of Weyl reflections of simply--laced root systems,\footnote{ See, for example, proposition VI.\S.\,1.33 of \cite{BOURB:LIE}} a Coxeter factorized sequence of mutations corresponds to a very peculiar finite BPS--chamber $\mathscr{C}_{\Lambda}$:
\begin{equation}\label{goodspectrum}
\mathscr{C}_{\Lambda} \simeq \bigoplus_{\a \in A} \Delta(G_\a)
\end{equation}
where by $\Delta(G)$ is meant the set of roots of $G$. In other words such a chamber contains one hypermultiplet per \emph{positive} root of $G_\a$. For a choice of generators compatible with $\mathscr{C}_{\Lambda}$
\be
\Gamma \simeq \bigoplus_{\a\in A} \Gamma(G_\a),
\ee
where we denote with $\Gamma(G_\a)$ the root lattice of the Lie algebra of type $G_\a$. It is useful to remark that any quiver that admits in its mutation class a square product form of type $Q \square G$ admits Coxeter-factorized sequences of type
\be
\underbrace{G\coprod G \coprod \cdots \coprod G}_{\#(Q_0) \text{ times}}
\ee
Many examples of these sequences can be found in the next sections.

\subsection{2d/4D and properties of the quantum monodromy}\label{tbw}
The 2d/4D worldsheet/target correspondence of \cite{CNV} is the statement that any 4D $\cn=2$ model with BPS quiver corresponds to a 2d $\cn=(2,2)$ model with $\hat{c}_\text{uv}<2$ that has the same BPS quiver. The correspondence is motivated by the geometric engineering of these models in Type II B superstrings. Consider the 3 CY hypersurface of $\C^4$ defined by the equation
\be\label{SWgeom}
\mathscr{H} \colon W(X_1,X_2,X_3,X_4) = 0 \subset \C^4.
\ee
The function $W$ is interpreted as the superpotential of a LG 2d $(2,2)$ system. If $W$ is an isolated  quasi homogeneous singularity, namely if
\be\label{quasihom}
\l^r \cdot W(X_i) \equiv W(\l^{w_i} X_i) \qquad r,w_i \in \mathbb{N} \mid \text{gcd}[r,w_1,w_2,w_3,w_4]=1
\ee
the 2d model is conformal and the 3 CY is singular. The singularity is at finite distance in CY moduli space provided $\hat{c}<2$ \cite{GVW}, and in this case we obtain a 4D $\cn=2$ SCFT. Whenever $W$ is not quasi homogeneous and $\hat{c}_\text{uv}<2$, both the 4D and the 2d models are asymptotically free. Many of the non perturbative aspects are described by the corresponding geometry. A crucial r\^ole is played by the Calabi Yau homolorphic top form, the Poincar\'e residue
\be
\Omega = \text{Res} \left( \frac{d X_1 \wedge d X_2 \wedge d X_3 \wedge d X_4}{W(X_1,X_2,X_3,X_4)} \right),
\ee
known also as the Seiberg Witten form. Consider the case of a superconformal theory. In the LG case, the 2d chiral ring of chiral primary operators corresponds to the Jacobian ideal associated to the superpotential: 
\be
\mathscr{R} \equiv \C[X_1,X_2,X_3,X_4]/\langle\partial_i W\rangle
\ee
For an element $\Phi \in \mathscr{R}$, let us denote by $q_\Phi$ its 2d $U(1)_R$ charge. Notice that if we identify the scaling symmetry of $W$ with the $R$ symmetry of the 2d model, requiring that $R[W] = 1$ we obtain
\be
\begin{aligned}
&q(X_i) \equiv w_i/r\\
&\hat{c} \equiv 4 - 2 \sum w_i/r
\end{aligned}
\ee
In particular, the 2d quantum monodromy has order $\text{lcm}(2,r)$. The scaling symmetry $X_i \to \l^{w_i}X_i$ acts on $\Omega$ as follows
\be\label{womega}
\Omega \longrightarrow \l^{w_\Omega} \cdot\Omega \qquad w_\Omega \equiv -r + \sum_i w_i = (2-\hat{c})r/2
\ee
The 4D scaling dimensions are fixed by requiring that the holomorphic top form has dimension one. Hence the 4D monodromy corresponds to the transformation
\be
X_i \to \text{exp}\left( 2 \pi i \frac{w_i}{w_\Omega} \right) X_i
\ee
and the order $\ell$ of the 4D monodromy is
\be
\ell = \frac{w_\Omega}{\text{gcd}(w_\Omega,w_1,w_2,w_3,w_4)}
\ee
In the case $W(X_1,X_2,X_3,X_4) = w(X_1,X_2)+X_3^2+X_4^2$, the order reduces, corresponding to the fact that the model admits an engineering in terms of the 6d $(2,0)$ theory. The reduced order is obtained from the above formula replacing $w_3 = d$, $w_4=0$. The various chiral primary deformations of $W$ define a family of 3 CY hypersurfaces:
\be
\mathscr{H}_{ \{ u_{\Phi} \} }  \colon W + \sum_{\Phi\in\mathscr{R}} u_{\Phi} \Phi = 0 \qquad u_{\Phi} \in \C
\ee
Let us denote by $\Omega_{\{u_{\Phi}\}}$ the corresponding family of Seiberg Witten forms. Deforming the CY geometry induce a massive deformation of corresponding 4D SCFT with primary operators $\co_{\Phi}$ with dual parameters $u_{\Phi}$. The scaling dimensions of the $\co_\Phi$ are determined by the requirement that $D[\co_{\Phi}] = 2 - D[u_\Phi]$, and the dimensions $D[u_\Phi]$ are fixed requiring that $D[\Omega_{\{u_{\Phi}\}}] = 1$, namely
\be
D[\Omega_{\{u_{\Phi}\}}] \equiv 1 \Longrightarrow\begin{cases} D[\Phi] = w_\Phi/w_\Omega \\D[u_{\Phi}] = (1-w_\Phi)/w_\Omega\end{cases}
\ee
The elements of the family $\mathscr{H}_{ \{ u_{\Phi} \} }$ that corresponds to physical deformations that parametrizes the IR Coulomb branch of the system are in correspondence with the $u_\Phi$ such that $D[u_{\Phi}] > 1$. From now on let us consider only these deformations of the geometry. The BPS particles are in correspondence with the vanishing special lagrangian 3 cycles $\arg (\Omega_{\{u_{\Phi}\}})|_L = \text{const}$. Given a basis of such 3 cycles $\{L_i\}$ in $H^3(\mathscr{H}_{\{u_\Phi\}},\bZ)$, it corresponds to a basis for the charge lattice $\{e_i\}$ with the BPS quiver property. The central charge is then captured by
\be\label{periods}
Z_i \equiv \int_{L_i} \Omega_{\{u_\Phi\}}
\ee
The BPS quiver computed out of the $tt^*$ Stokes matrix $S$ of the corresponding 2d model
\be
B \equiv S^t - S
\ee
coincides with the intersection pairing in between the vanishing 3 cycles (and mutation gets related with the usual Picard--Lefschetz transformations). Let $\ca_{\{u_\Phi\}}$ denote the set of automorphisms of the geometry $\mathscr{H}_{\{u_\Phi\}}$. Let  $\ca_0$ denote the subgroup of $\ca_{\{u_\Phi\}}$ that acts trivially on $\Omega_{\{u_\Phi\}}$. The BPS chamber corresponding to the stability condition encoded in eqn.\eqref{periods} has a $1/m$ fractional monodromy precisely when there is a deformations $\mathscr{H}_{\{u_\Phi\}}$ such that $\text{Card } \ca_{\{u_\Phi\}} / \ca_0  = m \in 2\mathbb{Z}_{>1}$. Notice that for generic values of the deformations $\ca_{\{u_\Phi\}} / \ca_0=\mathbb{Z}_2$, in correspondence with \textsc{pct} symmetry (orientation reversal). However, due to the wall crossing invariance of $\mathscr{Y}_q$, if $\mathscr{M}_q = \mathscr{Y}_q ^m$ in one chamber (\textit{E.g.}\! for one specific value of the deformations $\{u^*_\Phi\}$), the same factorization holds everywhere in the moduli. In particular, if the LG is quasi homogeneous, it is sufficient to choose the deformation of $W$ induced by the identity operator of $\mathscr{R}$, to see that the finest $1/m$ fractional monodromy has order equal to the 2d monodromy
\be
m \equiv \text{lcm}(2,r).
\ee

\section{The $Y$ system of a 4D $\cn=2$ model}

\subsection{From $\mathscr{Y}_q$ to $Y_{i,s}$}

The action of a quantum mutation on $\bT_\Gamma(q)$ is described in eqn.\eqref{explicitquantumut}. Let us consider the classical limit $q\to 1$ of eqn.\eqref{explicitquantumut}:
\be\label{classicalY}
\cq_i \cdot Y_j \Big|_{q\to1}=\begin{cases}
Y_i^{-1} & \text{if } i = j\\
 Y_j \,(1+ Y_i^{-1})^{- B_{ij}} &\text{if } B_{ij}\geq 0\\
Y_j \, (1+Y_i)^{|B_{ij}|}  &\text{if } B_{ij}\leq 0\\
\end{cases}
\ee
That is precisely the $Y$ seed mutation of \cite{kellerper}. The classical limit of a quantum mutation $\cq_i$ defines $n$ rational functions in the $n$ variables $\{Y_i\}$,
\be
R^{(i)}_k(Y_1,\dots,Y_n) \equiv \cq_i \cdot Y_k \big|_{q\to1} \qquad k = 1,...,n.
\ee
In particular, if the quiver is simply laced, \emph{i.e.} if $|B_{ij}|\leq 1$, 
\be
\cq_i \cdot Y_j = N[R^{(i)}_j(Y_1,\dots,Y_n)].
\ee

\medskip

Consider a 4D $\cn=2$ model with BPS quiver that have a finite chamber made only of hypermultiplets. Let $\Xi \equiv \{k_1,\dots, k_s\}$ be the sequence of nodes associated to its finer $1/m$ fractional quantum monodromy via the mutation method discussed in section \ref{WC}. The classical limit of the quantum mutation seqence corresponding to the fractional monodromy $\mathscr{Y}_q$ is a rational map
\be
R\colon\C^n\rightarrow \C^n \qquad Y_j\mapsto R_j(Y_1,\dots,Y_n)
\ee
where $R_j$ is given by
\be\label{ratmap}
R_j(Y_1,\dots,Y_n)\equiv \mathscr{I_\sigma}^{-1} \circ R^{(k_s)}_j \circ R^{(k_{s-1})}_j\circ\cdots \circ R^{(k_2)}_j\circ R^{(k_1)}_j (Y_1,\dots,Y_n)
\ee
The $Y$ system associated to a finite chamber of a $\cn=2$ model is defined as the recursion relation generated by the iteration of the rational map \eqref{ratmap}, namely
\begin{equation}\label{Ys}
 Y_{j,s+1} \equiv R_j\big(Y_{1,s},\dots,Y_{n,s}\big),\qquad s\in \Z.
\end{equation}

\medskip

\noindent{\bf Log symplectic property.} The rational map corresponding to the full quantum monodromy
\be
M_i(Y_1,\dots,Y_n) \equiv R_i \circ R_i \circ \cdots \circ R_i (Y_1,\dots,Y_n),
\ee
where $R_i$ is composed $m$ times, has the log--symplectic property
\begin{equation}
\langle e_i \,, e_j \rangle\: d\log Y_i\wedge d\log Y_j =  
\langle e_i\,, e_j\rangle\: d\log M_i\wedge d\log M_j.  
\end{equation}
Notice that for simply laced quivers, the action of $\mathscr{M}_q$ on the quantum torus algebra is fully determined by its classical counterpart:\begin{equation}
\textrm{Ad}^\prime(\mathscr{M}_q) \cdot Y_j \equiv N[M_j (Y_1,...,Y_n)].
\end{equation}
In particular, the periodicity of the latter, entails the periodicity of the former: if the quiver is simply laced the periodicity of the quantum $Y$ system follows from the periodicity of the classical one.

\subsection{SCFT and periodic $Y$ systems.} 

The above construction shows that the quantum monodromy inner automorphism of $\mathbb{T}_\Gamma(q)$ has finite order $\ell$ iff the rational map of eqn.\eqref{ratmap} has order $m \times \ell$, \emph{i.e} the corresponding $Y$ system is periodic with period $m\times \ell$. As we have remarked above, in the simply laced case one can show that also the converse is true, that is:
\begin{equation}
 \mathrm{Ad}^\prime(\mathscr{M}_q)^{\ell}=\mathrm{Id}_{\bT_\Gamma}\quad\Longleftrightarrow\quad Y_{j,s+m \ell}=Y_{j,s},\ \:\forall\, j\in Q_0,\:s\in \Z.
\end{equation}

\medskip

\noindent{\bf Remark.} It might happen that the smallest period of the $Y$ system actually is $k\times \ell$ for a positive integer $k\mid m$. Examples of this reduction are presented below, as well as an example, the $Y$ system associated to $D_2(SU(5))$, in which the minimal period is precisely $m \times \ell$ and no reduction occurs.

\medskip

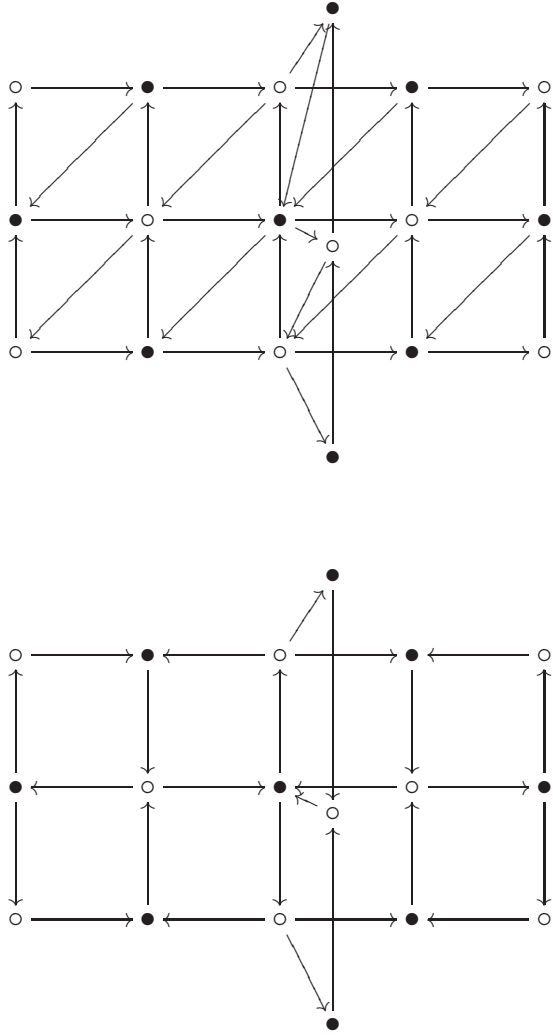
\begin{figure}
$$ 
\begin{xy} 0;<1pt,0pt>:<0pt,-1pt>:: 
(0,150) *+{\circ} ="0",
(50,150) *+{\bullet} ="1",
(100,150) *+{\circ} ="2",
(150,150) *+{\bullet} ="3",
(200,150) *+{\circ} ="4",
(120,190) *+{\bullet} ="5",
(0,100) *+{\bullet} ="6",
(50,100) *+{\circ} ="7",
(100,100) *+{\bullet} ="8",
(150,100) *+{\circ} ="9",
(200,100) *+{\bullet} ="10",
(120,110) *+{\circ} ="11",
(0,50) *+{\circ} ="12",
(50,50) *+{\bullet} ="13",
(100,50) *+{\circ} ="14",
(150,50) *+{\bullet} ="15",
(200,50) *+{\circ} ="16",
(120,20) *+{\bullet} ="17",
"0", {\ar"1"},
"0", {\ar"6"},
"7", {\ar"0"},
"1", {\ar"2"},
"1", {\ar"7"},
"8", {\ar"1"},
"2", {\ar"3"},
"2", {\ar"5"},
"2", {\ar"8"},
"9", {\ar"2"},
"11", {\ar"2"},
"3", {\ar"4"},
"3", {\ar"9"},
"10", {\ar"3"},
"4", {\ar"10"},
"5", {\ar"11"},
"6", {\ar"7"},
"6", {\ar"12"},
"13", {\ar"6"},
"7", {\ar"8"},
"7", {\ar"13"},
"14", {\ar"7"},
"8", {\ar"9"},
"8", {\ar"11"},
"8", {\ar"14"},
"15", {\ar"8"},
"17", {\ar"8"},
"9", {\ar"10"},
"9", {\ar"15"},
"16", {\ar"9"},
"10", {\ar"16"},
"11", {\ar"17"},
"12", {\ar"13"},
"13", {\ar"14"},
"14", {\ar"15"},
"14", {\ar"17"},
"15", {\ar"16"},
\end{xy}
$$

$$
\begin{xy} 0;<1pt,0pt>:<0pt,-1pt>:: 
(0,150) *+{\circ} ="0",
(50,150) *+{\bullet} ="1",
(100,150) *+{\circ} ="2",
(150,150) *+{\bullet} ="3",
(200,150) *+{\circ} ="4",
(120,190) *+{\bullet} ="5",
(0,100) *+{\bullet} ="6",
(50,100) *+{\circ} ="7",
(100,100) *+{\bullet} ="8",
(150,100) *+{\circ} ="9",
(200,100) *+{\bullet} ="10",
(120,110) *+{\circ} ="11",
(0,50) *+{\circ} ="12",
(50,50) *+{\bullet} ="13",
(100,50) *+{\circ} ="14",
(150,50) *+{\bullet} ="15",
(200,50) *+{\circ} ="16",
(120,20) *+{\bullet} ="17",
"0", {\ar"1"},
"6", {\ar"0"},
"2", {\ar"1"},
"1", {\ar"7"},
"2", {\ar"3"},
"2", {\ar"5"},
"8", {\ar"2"},
"4", {\ar"3"},
"3", {\ar"9"},
"10", {\ar"4"},
"5", {\ar"11"},
"7", {\ar"6"},
"6", {\ar"12"},
"7", {\ar"8"},
"13", {\ar"7"},
"9", {\ar"8"},
"11", {\ar"8"},
"8", {\ar"14"},
"9", {\ar"10"},
"15", {\ar"9"},
"10", {\ar"16"},
"17", {\ar"11"},
"12", {\ar"13"},
"14", {\ar"13"},
"14", {\ar"15"},
"14", {\ar"17"},
"16", {\ar"15"},
\end{xy}
$$
\caption{UP: The quiver $\vec{A}_3 \, \boxtimes \, \vec{E}_6$. DOWN: The quiver $A_3 \, \square \, E_6$. 
The Coxeter factorized sequence of mutation of type $E_6 \coprod E_6 \coprod E_6$ is simply obtained mutating 
first all $\circ$'s then all $\bullet$'s and iterating. The one of type $A_3\coprod\dots\coprod A_3$ mutating 
first all $\bullet$'s then all $\circ$'s and iterating.}\label{square}
\end{figure}

\noindent{\bf Zamolodchikov's $Y$ systems.} The definition of the $Y$ system in eqn.\eqref{Ys} is motivated
 by the reconstruction of the Zamolodchikov TBA periodic $Y$ systems for the integrable 2d $(G,G^\prime)$ models.
 The proof of the periodicity of these systems was achieved using precisely the
 formalism of cluster algebras in \cite{CNV,kellerper}. In our setup, periodicity of the $Y$ system is predicted 
by the physical properties of a model, and, in a sense, explained in terms of superconformal symmetry.
 As an interesting motivating example, let us review here the details of the original construction of \cite{CNV}. 

\medskip

The ADE singularities are listed in table \ref{ADEsings} together with some of their fundamental properties.
 The 4D $\cn=2$ SCFT's of type $(G,G^\prime)$ is obtained, for example, by geometric  
engineering the Type II B superstring on the direct sum singular hypersurface of $\C^4$
\be\label{GGpot}
W_G(X_1,X_2)+W_{G^\prime}(X_3,X_4) = 0.
\ee
By 2d/4D worldsheet/target correspondence the LHS of eqn.\eqref{GGpot} can be interpreted as
 the superpotential of a 2d $\cn=(2,2)$ SCFT. The 2d model is the direct sum of two 
\emph{non interacting} 2d SCFT's: the Hilbert space of the system is simply the tensor product 
of the two Hilbert spaces of the factors. The central charge $\hat{c}$ of the $(G,G^\prime)$ 2d model is
\be
\hat{c} \equiv \hat{c}_G + \hat{c}_{G^\prime} < 2.
\ee
Recall that the $tt^*$ Stokes matrix of the minimal LG model of type $G$ is the upper triangular matrix with unit diagonal
\be\label{Gquiv}
(S_G)_{ij} \equiv \d_{ij} + \begin{cases} (C_G)_{ij} &\text{if }i < j  \\
0 &\text{else}\end{cases}
\ee
The quiver one obtains is:
\be
\vec{G}\equiv S_G^t - S_G.
\ee
The quiver $\vec{G}$ is obtained from the Dynkin graph of type $G$ by giving the same orientation to all edges. Since the Hilbert space of the 2d $(G,G^\prime)$ model is a tensor product $\mathcal{H}_G\otimes \mathcal{H}_{G^\prime}$, the $tt^*$ Stokes matrix is simply
\be
S \equiv S_G \otimes S_{G^\prime}.
\ee
The BPS quiver of the system is then obtained by 2d/4D correspondence. It is encoded by
\be\label{triangleprod}
B \equiv (S_G\otimes S_{G^\prime})^t - S_G\otimes S_{G^\prime}.
\ee
The quiver (with superpotential) associated to this intersection matrix is the triangular tensor product quiver $\vec{G} \, \boxtimes \, \vec{G}^\prime$, an interesting element of the quiver mutation class associated to the systems of type $(G,G^\prime)$. For an example see figure \ref{square}. However, this is not the representative we are after. By a small abuse of notation, let us denote with the letter $G$ the quiver obtained giving to the simply laced Dynkin graph of type $G$ an alternating orientation, \textit{i.e.}\! each vertex of $G$ is a source or a sink. By mutating only on sinks of the $G$ vertical subquivers and on sources of the $G^\prime$ ones, it is very easy to construct mutation sequences that connects the quiver $\vec{G} \, \boxtimes \, \vec{G}^\prime$ to the square tensor product quiver $G \, \square \, G^\prime$. The latter is obtained from $\vec{G} \, \boxtimes \, \vec{G}^\prime$ by setting to zero all the `diagonal' arrows, replacing all $\vec{G}$ vertical subquivers and all $\vec{G}^\prime$ horizontal subquivers with their alternating versions $G$ and $G^\prime$, and reversing all arrows in the full subquivers of the form $\{i\}\times G^\prime$ and $G\times \{i^\prime\}$, where $i$ is a sink of $G$ and $i^\prime$ a source of $G^\prime$. For an example see figure $\ref{square}$. The latter is clearly the better option for searching Coxeter factorized sequences of mutations. 

\begin{table}
\begin{center}
\begin{tabular}{|c|c|c|c|c|}
\hline
& $W_G(X,Y)$ & $(q_X,q_Y)$ & $\hat{c}$ &$h(G)$\\
\hline
$A_{n-1}$ & $X^n + Y^2$ & $(1/n,1/2)$ & $\frac{n-2}{n}$\phantom{$\Bigg|$}&$n$\\
\hline
$D_{n+1}$ & $X^n + XY^2$ &$ (1/n, (n-1)/2n)$ & $\frac{(2-n^2)(n-1)}{n}$\phantom{$\Bigg|$}&$2n$ \\
\hline
$E_6$ & $X^3 + Y^4 $& $(1/3,1/4)$ & $\frac{5}{6}$\phantom{$\Bigg|$}&12 \\
\hline
$E_7$ & $X^3 + X Y ^3 $& $(1/3,2/9)$ &$\frac{8}{9}$\phantom{$\Bigg|$} &18\\
\hline
$E_8$ & $X^3 + Y^5 $& $(1/3, 1/5)$ &$\frac{14}{15}$ \phantom{$\Bigg|$}&30 \\
\hline
\end{tabular}
\end{center}
\caption{ List of the $ADE$ simple singularities, and some of the corresponding properties. }\label{ADEsings}
\end{table}

\medskip

Let us label the nodes as follows: black nodes $\bullet$ are sources (resp.\! sinks) of the vertical $G$ (resp.\! horizontal $G^\prime$) subquivers, white nodes $\circ$ are sinks (resp.\! sources) of the vertical $G$ (resp.\! horizontal $G^\prime$) subquivers. Clearly the $(G,G^\prime)$ model admit fractional monodromies associated to Coxeter factorized sequences of mutations. Let us define
\be
\Xi_\bullet \equiv \{ \text{sequence of all } \bullet_i \text{ nodes}\} \qquad \Xi_\circ \equiv \{ \text{sequence of all } \circ_j \text{ nodes}\}
\ee
The ordering in the sub--sequences $\{\bullet_i\}$ and $\{\circ_j\}$ is irrelevant: since, by construction, distinct $\bullet_i$ nodes (resp. $\circ_j$ nodes) are not connected in the quiver $G \, \square \, G^\prime$, the corresponding elementary mutations commutes, \textit{e.g.}  $\mu_{\bullet_a}\mu_{\bullet_b} = \mu_{\bullet_b}\mu_{\bullet_a}$. Then we define
\be
\Xi \equiv \{\Xi_\bullet,\Xi_\circ\} \qquad\qquad \Xi^\prime \equiv \{\Xi_\circ,\Xi_\bullet\}.
\ee
Notice that $\Xi$ (resp. $\Xi^\prime$) correspond to the action of the Coxeter element $c$ of $\text{Weyl}(G)$  (resp. $\text{Weyl}(G^\prime)$) on the charge lattice $\Gamma$, and the corresponding finite BPS chambers are of type
\be
\begin{gathered}
G\underbrace{\coprod G \coprod \cdots\coprod}_{\text{rank } G^\prime \text{ times }}  G\\
\end{gathered}\quad\text{and}\quad \begin{gathered} G^\prime\underbrace{\coprod G^\prime \coprod \cdots\coprod}_{\text{rank } G \text{ times }}  G^\prime\end{gathered}\ee
The BPS chambers we have just constructed correspond via eqns.\eqref{ratmap}--\eqref{Ys} to the Zamolodchikov $Y$ system of type $(G,G^\prime)$.
 More precisely, the Zamolochikov's $Y$ system of eqn.\eqref{pairofG} 
corresponds to the rational maps associated with the sequences $\Xi_\bullet$ and $\Xi_\circ$. One has
\be
Y_{i,i^\prime,n+1} \equiv \begin{cases} R_{\Xi_\bullet}(\{ Y_{i,i^\prime,n}\}) &n \text{ odd}\\
R_{\Xi_\circ}(\{ Y_{i,i^\prime,n}\})&n \text{ even}
\end{cases}
\ee
Thus a shift $s \to s+1$ in the discrete time of the $Y$--system associated with the fractional monodromy we have constructed corresponds to a shift $n\to n+2$ in the original discrete time of the Zamolodchikov $Y$ systems. This is true with a small \emph{caveat}. In favorable circumstances it might happen that $\Xi_\bullet$ (or, equivalently, $\Xi_\circ$) itself corresponds to a fractional monodromy. In that case, there is a redundancy in the corresponding $Y$ system that reduces precisely the Zamolodchikov's one. Let us see an instance of this phenomenon in the following example:

\medskip

\noindent{\bf Example 1: The $Y$ system of type $(A_1,A_2)$.} The quiver is $A_1 \square A_2 \simeq A_2$. Set
\be
\bullet \longrightarrow \circ
\ee
In this case $\Xi_\circ$ is clearly the fractional monodromy, associated with the permutation $(\circ,\bullet)$. We obtain
\be
\begin{cases}
Y_{\bullet,s+1} = Y_{\circ,s}^{-1}\\
Y_{\circ,s+1} = Y_{\bullet,s} (1+Y_{\circ,s}).
\end{cases}
\ee
Eliminating the redundant $Y_\bullet$ we are left with the Zamolodchikov Y-system of $A_2$ type:
\be
Y_{\circ,s+1}Y_{\circ,s-1} = 1+Y_{\circ,s}.
\ee

\medskip

\noindent{\bf Remark.} We can anticipate that this type of reduction will be possible for all those pair of $G$ and $G^\prime$ that are such that
\be
(G \, \square \, G^\prime)^\text{op} = \sigma (G \, \square \, G^\prime),
\ee
where $\sigma$ is a permutation of the labels of the nodes. Indeed, both sequences of mutations $\mathbf{m}_{\Xi_\circ}$ and $\mathbf{m}_{\Xi_\bullet}$ map $G \, \square \, G^\prime$ to $(G \, \square \, G^\prime)^\text{op}$.

\medskip

We stress here that the sequences of mutations $\Xi$ and $\Xi^\prime$ \emph{solve} the $Y$ system recursion.

\medskip

\noindent{\bf Example 2: The $Y$ system of type $(A_1,A_3)$.} Let us consider the next case: $A_3 \simeq A_1 \square A_3$. In this case, nor $\Xi_\bullet$ nor $\Xi_\circ$ have the right properties for being interpreted as fractional monodromies, and the minimal choice is given by either $\Xi$ or $\Xi^\prime$. Let us label the nodes as follows:  
\be
\bullet_1 \longrightarrow \circ \longleftarrow \bullet_2
\ee
The $Y$ system associated with the fractional monodromy $\Xi=\{\bullet_1,\bullet_2,\circ\}$, for example, is
\bea
Y_{\bullet_1,s+1} &= \frac{Y_{\circ,s} Y_{\bullet_2,s}}{1 + Y_{\bullet_1,s} + (1 + Y_{\bullet_1,s} + Y_{\bullet_1,s}Y_{\circ,s})Y_{\bullet_2,s})}\\
Y_{\circ,s+1} &=\frac{(1 + Y_{\bullet_1,s})(1 + Y_{\bullet_2,s})}{Y_{\bullet_1,s}Y_{\circ,s}Y_{\bullet_2,s}}\\
Y_{\bullet_2,s+1} &= \frac{Y_{\bullet_1,s}Y_{\circ,s}}{1 + Y_{\bullet_1,s} + (1 + Y_{\bullet_1,s} + Y_{\bullet_1,s}Y_{\circ,s})Y_{\bullet_2,s}}\eea
And one can easily check that the above has period 6.

\medskip

\noindent{\bf Remark.} Besides the models of class $(G,G^\prime)$, many other models with $R$ charges in $\mathbb{N}/\ell$ and finite BPS chambers of hypers can be constructed from singularity theory along the lines of section \ref{tbw}. In particular in \cite{Arnold1,Arnold2} the models corresponding to unimodal and bimodal Arnol'd singularities have been analyzed according to the principles we have discussed above. In appendix \ref{arn} we report the relevant tables from \cite{Arnold1, Arnold2}.

\medskip

\noindent{\bf Remark 2.} It would be interesting to give an interpretation of the new periodic $Y$ systems 
corresponding to $\cn=2$ SCFT's in terms of exactly solvable $2d$ theories in analogy with the $(G,G^\prime)$ 
ones \cite{Zamo,tateo,kuniba,volkov}. 

\medskip

\noindent{\bf Unexpected relation with Cremona groups.} At the mathematical level, we get an unexpected relation
 between singularity theory and cyclic subgroups of the Cremona groups $\mathrm{Cr}(n)$ of birational automorphisms
 $\mathbb{P}^n\rightarrow \mathbb{P}^n$, both interesting subjects in Algebraic Geometry (the second one being notoriously
 hard for $n\geq 3$ \cite{dolgachev}). We stress that, although the explicit form of the $Y$
 system depends on the particular finite BPS chamber we use to write the map \eqref{ratmap}, two $Y$ systems
 corresponding to different chambers of the \emph{same} $\cn=2$ theory are equivalent, in the sense that they are 
related by a rational change of variables $Y_j\rightarrow Y^\prime_j(Y_k)$. Indeed, the monodromies are independent of 
the chambers up to conjugacy, and so is its classical limit map $Y_j\rightarrow R_j$. Hence the rational maps $R_j$ obtained
 in different chambers are conjugate in the Cremona group.

\subsection{A detailed example: Minahan--Nemeschansky theories}
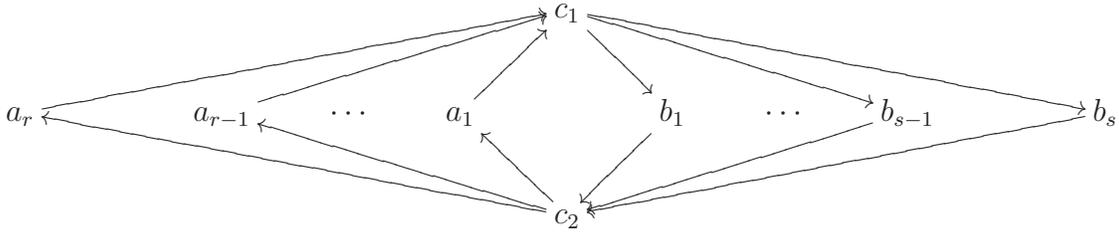
\begin{figure}
$$
\xymatrix{
& & & & &c_1 \ar[dr]\ar[drrr]\ar[drrrrr] &&&&\\ 
a_r\ar[urrrrr]&&a_{r-1}\ar[urrr]&\cdots&a_1\ar[ur]&&b_1\ar[dl]&\cdots&b_{s-1}\ar[dlll] && b_s\ar[dlllll] \\
& & & & &c_2 \ar[ul]\ar[ulll]\ar[ulllll] &&&& }
$$
\caption{The quivers in the family $Q(r,s)$.}\label{MNquivs}
\end{figure}
The 4D $\cn=2$ rank 1 SCFT's in correspondence with Kodaira's list of singular fibers have specially simple BPS quivers, $Q(r,s)$ drawn in figure \ref{MNquivs}. We have the following identification:
\be\label{MNidqu}
\begin{tabular}{|c|c|c|c|c|c|c|}
\hline
SCFT&$H_2$&$H_3$&$D_4$&$E_6$&$E_7$&$E_8$\\
\hline
quiver&$Q(1,0)$&$Q(1,1)$&$Q(2,2)$&$Q(3,3)$&$Q(4,3)$&$Q(5,3)$\\
\hline
\end{tabular}
\ee
This can be motivated as follows: the flavor group $F$ of any of the above $\cn=2$ rank 1 SCFT is a star shaped Dynkin with arms of length $[r,s,2]$. We assign the BPS quiver $Q(r,s)$ to the model with flavor group corresponding to the Dynkin graph $[r,s,2]$.\footnote{ For example $D_4$ is the star $[2,2,2]$ and corresponds to $Q(2,2)$.} The models $H_2$ and $H_3$ corresponds to the Argyres--Douglas theories of type $A_3$ and $D_4$. The model labeled by the $D_4$ fiber corresponds to the lagrangian superconformal theory $SU(2)$ $N_f = 4$. The models labeled by exceptional flavor groups, instead, corresponds to the strongly interacting non lagrangian exceptional  Minahan--Nemeschansky SCFT's. The two Argyres--Douglas models above are mapped to the Zamolodchikov's $Y$ systems of type $(A_3,A_1)$ and $(D_4,A_1)$. Here we focus on the $Y$ systems that are not of Zamolodchikov type.

\medskip

\noindent{\bf Fractional monodromies and SW curves.} The SW curves for these systems (turning off all possible mass deformations) are the following elliptic curves
\be\label{rank1}
\begin{tabular}{ll}
$E_8\colon y^2 = x^3 + 2 u^5 $&$E_6 \colon y^2=x^3+u^4 $\\ 
$E_7 \colon y^2=x^3 + u^3\, x $&$D_4 \colon	y^2 = x^3 + 3\tau \, u^2\, x + 2 u^3 $\\
\end{tabular}
\ee
Where $u$ denotes the Coulomb branch parameter, and for the case of $D_4$, $\tau$ is the exactly marginal UV $SU(2)$ coupling. The Seiberg Witten differential in this case satisfies 
\be
\partial_u \l_{\text{SW}}  = \frac{d x} { y} + \frac{d}{dx}(\text{ somewhat })
\ee 
According to our general principles outlined in section \S.\ref{tbw}, deforming eqn.\eqref{rank1} with a constant mass term to flow in the IR Coulomb branch, the discrete subgroup of the group of autmorphisms of the SW curves that acts non trivially on $\l_{\text{SW}}$ is
\be\label{rank1}
\begin{tabular}{ll}
$E_8\colon \bZ_2 \times \bZ_3 \times \bZ_5 $&$E_6 \colon \bZ_3 \times \bZ_4$\\ 
$E_7 \colon\bZ_2 \times \bZ_9$&$D_4 \colon \bZ_2 \times \bZ_3$\\
\end{tabular}
\ee
All these groups have order $h(F)$. Therefore we predict these systems have a $1/h(F)$ fractional monodromy. Since these models have quantum monodromies of order $\ell =1$, the period of the corresponding $Y$ systems is $h(F)$.

\medskip

\noindent{\bf  The $SU(2)$ $N_f=4$ $Y$ system.} Let us relabel the nodes of the quiver $Q(2,2)$ as follows:
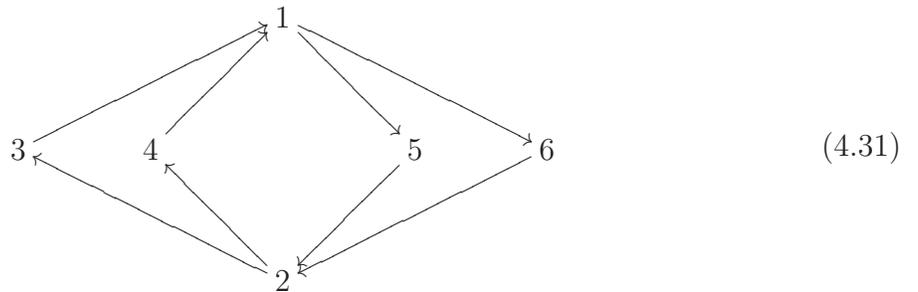
\be
\begin{gathered}\begin{xy} 0;<1pt,0pt>:<0pt,-1pt>:: 
(100,0) *+{1} ="0",
(100,100) *+{2} ="1",
(0,50) *+{3} ="2",
(50,50) *+{4} ="3",
(150,50) *+{5} ="4",
(200,50) *+{6} ="5",
"2", {\ar"0"},
"3", {\ar"0"},
"0", {\ar"4"},
"0", {\ar"5"},
"1", {\ar"2"},
"1", {\ar"3"},
"4", {\ar"1"},
"5", {\ar"1"},
\end{xy}
\end{gathered}
\ee
With this notation we have
\be
\Xi \equiv \{1,2,3,5\} \qquad\sigma \equiv \{(1,2),(3,4,5,6)\}
\ee
This is a sequence associated to a fractional monodromy because
\be
\mathbf{m}_{\sigma^2(\Xi)}\circ\mathbf{m}_{\sigma(\Xi)}\circ\mathbf{m}_{\Xi}(e_i) = -e_{\pi(i)}.
\ee
Where, of course,
\be
\pi = \{(1,2),(3,4,5,6)\}.
\ee
This shows that the mutation sequence we have found is a $1/6$ fractional monodromy. The corresponding $Y$ system is encoded by the following birational map $\C^6 \to \C^6$
\be
\begin{aligned}
&Y_{1,s+1} = \frac{Y_{3,s} (1 + Y_{1,s} + Y_{1,s} (1 + Y_{2,s}) Y_{5,s})}{1 + Y_{2,s} + (1 + Y_{1,s}) Y_{2,s} Y_{3,s}} & Y_{4,s+1} = \frac{1 + Y_{1,s}}{Y_{1,s} Y_{5,s} + Y_{1,s} Y_{2,s} Y_{5,s}}\\
&Y_{2,s+1} = \frac{(1 +  Y_{2,s} + (1 + Y_{1,s}) Y_{2,s} Y_{3,s}) Y_{5,s}}{1 + Y_{1,s} +  Y_{1,s} (1 + Y_{2,s}) Y_{5,s}} & Y_{5,s+1} = \frac{Y_{1,s} (1 + Y_{2,s}) Y_{6,s}}{1 + Y_{1,s}}\\
&Y_{3,s+1} = \frac{(1 + Y_{1,s}) Y_{2,s} Y_{4,s}}{1 + Y_{2,s}} & Y_{6,s+1} =  \frac{1 + Y_{2,s}}{Y_{2,s} Y_{3,s} + Y_{1,s} Y_{2,s} Y_{3,s}}\\
\end{aligned}
\ee
One can easily check numerically that the above map is periodic of period 6: $Y_{i,s+6} \equiv Y_{i,s}$.

\medskip

\noindent{\bf The other $Y$ systems.} The sequence of nodes that corresponds to the minimal fractional monodromy for the quivers $Q(3,3)$, $Q(4,3)$, and $Q(5,3)$ have always the form
\be\label{MNseq}
\Xi \equiv \{c_1,c_2,a_1,b_1\}
\ee
The mutation sequence $\mathbf{m}_\Xi$ acts as the identity on $Q(r,s)$ up to a cyclic permutation of its nodes of order $\text{lcm}(2,r+s)$, namely
\be\label{MNperm}
\sigma \equiv \{(c_1,c_2),(a_1,a_2,\dots,a_{r-1},a_r,b_1,b_2,\dots,b_{s-1},b_s)\}\in \mathfrak{S}_{2+r+s}
\ee
The explicit form of the $Y$ system recursion $\C^n \to \C^n$ for the pairs $(r,s) = (3,3),(4,3),(5,3)$ takes the form
\bea
Y_{c_1,n+1} &= \frac{(1+Y_{c_1,n}(1+(1+Y_{c_2,n}) Y_{b_1,n}))Y_{a_1,n}}{1+Y_{c_2,n}(1+(1+Y_{c_1,n})Y_{a_1,n})}\\
Y_{c_2,n+1} &= \frac{1+Y_{c_2,n}(1+(1+Y_{c_1,n})Y_{a_1,n})Y_{b_1,n}}{(1+Y_{c_1,n}(1+(1+Y_{c_2,n}) Y_{b_1,n}))}\\
Y_{a_i,n+1}&= \frac{(1+Y_{c_1,n})Y_{c_2,n} Y_{a_{i+1,n}}}{1+Y_{c_2,n}}\qquad i=1,...,r-1\\
Y_{b_j,n+1}&=\frac{(1+Y_{c_2,n}) Y_{c_1,n}Y_{b_{j+1},n}}{1+Y_{c_1,n}}\qquad j = 1,...,s-1\\
Y_{a_r,n+1}&= \frac{1+Y_{c_1,n}}{(1+Y_{c_2,n}) Y_{c_1,n}Y_{b_1,n}}\\
Y_{b_s,n+1}&=\frac{1+Y_{c_2,n}}{(1+Y_{c_1,n})Y_{c_2,n} Y_{a_{1,n}}}.
\eea

We have checked numerically the desired minimal periodicity of the above expressions is precisely $h(F)$.

\subsection{$Y$ systems of $D_2(G)$ theories.}
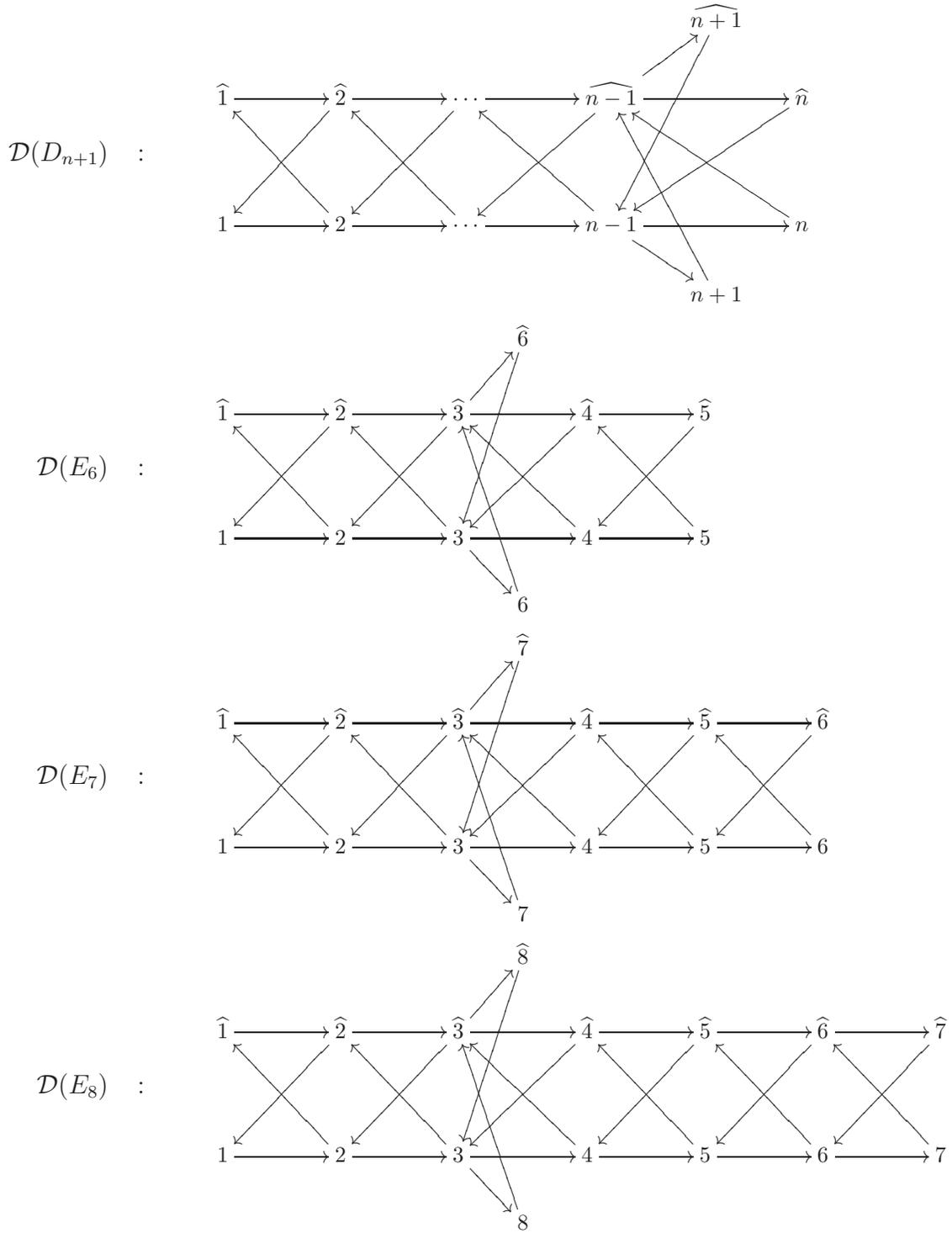
\begin{figure}
$$
\begin{aligned}
\cd(D_{n+1}) \quad\colon\qquad & {\footnotesize\begin{gathered}
\xymatrix@R=1.5pc@C=1.5pc{&&&&&&&\widehat{n+1}\ar[dddl]\\
\widehat{1}\ar[rr]&&\widehat{2}\ar[rr]\ar[ddll]&&\cdots\ar[ddll]\ar[rr]&&\widehat{n-1}\ar[ur]\ar[rr]\ar[ddll]&&\widehat{n}\ar[ddll]\\
&&\\
1\ar[rr]&&2\ar[rr]\ar[uull]&&\cdots\ar[rr]\ar[uull]&&n-1\ar[rr]\ar[dr]\ar[uull]&&n\ar[uull]\\
&&&&&&&n+1\ar[uuul]}
\end{gathered}}\\
\cd(E_6) \quad\colon\qquad &{\footnotesize \begin{gathered}
\xymatrix@R=1.5pc@C=1.5pc{&&&&&\widehat{6}\ar[dddl]\\
\widehat{1}\ar[rr]&&\widehat{2}\ar[rr]\ar[ddll]&&\widehat{3}\ar[ur]\ar[ddll]\ar[rr]&&\widehat{4}\ar[rr]\ar[ddll]&&\widehat{5}\ar[ddll]\\
&&\\
1\ar[rr]&&2\ar[rr]\ar[uull]&&3\ar[rr]\ar[uull]\ar[dr]&&4\ar[rr]\ar[uull]&&5\ar[uull]\\
&&&&&6\ar[uuul]}
\end{gathered}}\\
\cd(E_7) \quad\colon\qquad & {\footnotesize\begin{gathered}
\xymatrix@R=1.5pc@C=1.5pc{&&&&&\widehat{7}\ar[dddl]\\
\widehat{1}\ar[rr]&&\widehat{2}\ar[rr]\ar[ddll]&&\widehat{3}\ar[ur]\ar[ddll]\ar[rr]&&\widehat{4}\ar[rr]\ar[ddll]&&\widehat{5}\ar[rr]\ar[ddll]&&\widehat{6}\ar[ddll]\\
&&\\
1\ar[rr]&&2\ar[rr]\ar[uull]&&3\ar[rr]\ar[uull]\ar[dr]&&4\ar[rr]\ar[uull]&&5\ar[rr]\ar[uull]&&6\ar[uull]\\
&&&&&7\ar[uuul]}
\end{gathered}}\\
\cd(E_8) \quad\colon\qquad &{\footnotesize \begin{gathered}
\xymatrix@R=1.5pc@C=1.5pc{&&&&&\widehat{8}\ar[dddl]\\
\widehat{1}\ar[rr]&&\widehat{2}\ar[rr]\ar[ddll]&&\widehat{3}\ar[ur]\ar[ddll]\ar[rr]&&\widehat{4}\ar[rr]\ar[ddll]&&\widehat{5}\ar[rr]\ar[ddll]&&\widehat{6}\ar[rr]\ar[ddll]&&\widehat{7}\ar[ddll]\\
&&\\
1\ar[rr]&&2\ar[rr]\ar[uull]&&3\ar[rr]\ar[uull]\ar[dr]&&4\ar[rr]\ar[uull]&&5\ar[rr]\ar[uull]&&6\ar[rr]\ar[uull]&&7\ar[uull]\\
&&&&&8\ar[uuul]}
\end{gathered}}
\end{aligned}
$$
\caption{The DWZ--reduced quivers $\cd(G)$ for the $D_2(G)$ SCFT's.}\label{D2Gquivs}
\end{figure}

Another intriguing example of periodic $Y$ systems that are not of Zamolodchikov type is given by the theories of type $D_2(G)$. Even though a rigorous construction of these models is possible only via a categorical tinkertoy, one can treat them as if were engineered in type II B from the $Z \to \infty$ limit of the geometry
\be\label{d2geom}
e^{-Z} + e^{2 \, Z} + W_{G}(X,Y) + U^2 = 0.
\ee
Consider deforming eqn.\eqref{d2geom} with a constant term. The transformation 
\be
(X,Y,Z,U) \mapsto (\omega_X \,X , \omega_Y\,Y, Z + \text{log } \omega_Z , \omega_U\,U)
\ee
leaves the $Z \to \infty$ limit of eqn.\eqref{d2geom} invariant iff
\be
\omega_Z^2 = \omega_U^2 = 1 \text{ and } W_G(\omega_X \, X, \omega_Y \, Y) = W_G(X,Y). 
\ee
The SW holomorphic top form is mapped in
\be
\Omega \mapsto (\omega_X \, \omega_Y \,\omega_Z\,\omega_U) \cdot \Omega
\ee
A quick look at table \ref{ADEsings} is sufficient to establish that the system has a $1/m$ fractional monodromy of order $m=\text{lcm}(2,h)$. The BPS quiver of $D_2(G)$'s and their BPS spectrum was computed in \cite{CDZG}. Let us review here the details, and add the explicit expression of the corresponding $Y$ system. With the arguments of section \ref{tbw} applied to the chiral ring of the LG in eqn.\eqref{d2geom}, one can show that the quantum monodromy of these systems has order
\be\label{monordD2G}
\ell(2,G) \equiv \frac{2}{\text{gcd}(2,h(G))} 
\ee
where $h(G)$ is the Coxeter number of the ADE Dynkin graph $G$. The systems of type $D_2(G)$ have a DWZ reduced BPS quiver that we have denoted $\cd(G)$. For instance, the quiver $\cd(A_n)$ is
\begin{equation}\label{pqqq)I}
\cd(A_n)\equiv\begin{gathered}
\xymatrix{\widehat{1}\ar[rr]&&\widehat{2}\ar[rr]\ar[ddll]&&\cdots\ar[ddll]\ar[rr]&&\widehat{n-1}\ar[rr]\ar[ddll]&&\widehat{n}\ar[ddll]\\
&&\\
1\ar[rr]&&2\ar[rr]\ar[uull]&&\cdots\ar[rr]\ar[uull]&&n-1\ar[rr]\ar[uull]&&n\ar[uull]}
\end{gathered}
\end{equation}
The $\cd(G)$'s for the other simply laced Lie algebras being represented in figure \ref{D2Gquivs}.

\medskip

The quivers $\cd(G)$ contain two full Dynkin $G$ subquivers with alternating orientation and non--overlapping support. \textit{E.g.}\! the two alternating $A_n$ subquivers of $\cd(A_n)$ in \eqref{pqqq)I} are the full subquivers over the nodes
\be
\big\{1,2,\widehat{3},\widehat{4},5,6,\dots\big\}\quad \text{and}\quad \big\{\widehat{1},\widehat{2},3,4,\widehat{5},\widehat{6},\dots\big\}.
\ee
With reference to this example, let us define the following mutation sequence:
\be\label{mutD2}
{\bf m}_{2,A_n} \equiv \Bigg( \prod_{a \text{ even}} \mu_{a} \circ \mu_{\widehat{a}} \Bigg)\circ \Bigg( \prod_{a \text{ odd}} \mu_{a} \circ \mu_{\widehat{a}}\Bigg)
\ee
For systems with $h = n+1$ even, this is the seqence corresponding to the finest fractional monodromy compatible with the chamber $A_n \coprod A_n$. For systems with $h=n+1$ odd, the minimal fractional monodromy is associated to the odd part of eqn.\eqref{mutD2}. Indeed, the mutation sequence corresponding to the full quantum monodromy associated to the $A_n\oplus A_n$ chamber is symply
\be
({\bf m}_{2,A_n})^{n+1} = \underbrace{\ {\bf m}_{2,A_n} \circ \dots \circ {\bf m}_{2,A_n} \ }_{n+1 \text{ times}}.
\ee

\medskip

We draw the BPS-quivers of the other $D_2(G)$ models in figure \ref{D2Gquivs}. From their structure it is clear that they  all admit Coxeter--factorized sequences of type $G\coprod G$ constructed analogously. Then we expect that the finer $Y$ systems corresponding to the fractional monodromy of these systems have orders
\be\label{bestmonodr}
\text{lcm}(2,h) \times \ell(2,G) = 2 \times \frac{\text{lcm}(2,h)}{\text{gcd}(2,h)}.
\ee
We have numerically checked this prediction for the corresponding 2d solvable models getting perfect agreement. Let us give here an example.

\medskip

\noindent{\bf An explicit example: $D_2(SU(5))$}. Let us label the nodes of the BPS quiver for $\cd(SU(5))$ as follows
\be
\begin{gathered}
\begin{xy} 0;<1.2pt,0pt>:<0pt,-1.2pt>:: 
(0,50) *+{1} ="0",
(50,50) *+{2} ="1",
(100,50) *+{3} ="2",
(150,50) *+{4} ="3",
(0,0) *+{5} ="4",
(50,0) *+{6} ="5",
(100,0) *+{7} ="6",
(150,0) *+{8} ="7",
"0", {\ar"1"},
"5", {\ar"0"},
"1", {\ar"2"},
"1", {\ar"4"},
"6", {\ar"1"},
"2", {\ar"3"},
"2", {\ar"5"},
"7", {\ar"2"},
"3", {\ar"6"},
"4", {\ar"5"},
"5", {\ar"6"},
"6", {\ar"7"},
\end{xy}
\end{gathered}
\ee
We have chosen this model because it has a such a small  fractional monodromy that generating the corresponding $Y$ system is really trivial. Indeed, we have
\be
\Xi \equiv \{1,5,3,7\} \qquad \sigma \equiv \{ (1,4),(2,3),(5,8),(6,7) \}
\ee
The corresponding $Y$ system recursion is
\be
\begin{aligned}
&Y_{1,s+1}\equiv \frac{(1 + Y_{7,s})Y_{3,s} Y_{4,s} }{1 + Y_{3,s}}& Y_{2,s+1} \equiv \frac{1}{Y_{3,s}} \\
&Y_{3,s+1} \equiv \frac{Y_{1,s} Y_{2,s} Y_{7,s} (1 + Y_{3,s}) (1 + Y_{5,s})}{(1 + Y_{1,s}) (1 + Y_{7,s})} & Y_{4,s+1}\equiv \frac{1}{Y_{1,s}}\\ 
&Y_{5,s+1}\equiv \frac{(1 + Y_{3,s}) Y_{7,s} Y_{8,s}}{1 + Y_{7,s}} & Y_{6,s+1}\equiv \frac{1}{Y_{7,s}}\\
&Y_{7,s+1} \equiv \frac{Y_{3,s} Y_{5,s} Y_{6,s}(1 + Y_{1,s})(1 + Y_{7,s})}{(1 + Y_{3,s}) (1 + Y_{5,s})}&Y_{8,s+1}\equiv \frac{1}{Y_{5,s}}
\end{aligned}
\ee
And one can numerically check that its minimal period is $20 = 2 \times \text{lcm}(2,5)$:
\be
Y_{i,s+20} = Y_{i,s}
\ee
as predicted from eqn.\eqref{bestmonodr}.

\medskip

\noindent{\bf Remark.} Even more (new) periodic $Y$ systems can be obtained from the other systems of type $D_p(G)$ of \cite{infinitelymany,CDZG}.

\section{Asymptotically free models, friezes, and Q systems}\label{AF}
\subsection{Fractional monodromy and frieze pattern}
In the previous section we have discussed the $Y$ system canonically associated to the BPS spectral problem of 4D $\cn=2$ systems. In the special case of SCFT's such $Y$ system is periodic. The purpose of this section is to illustrate the special properties of the $Y$ system in case the model is asymptotically free. For this purpose, instead of working directly with it, is more convenient to  introduce the \emph{frieze pattern} $\{X_{i,s}\}$ with $i\in Q_0$, $s\in\mathbb{Z}$ associated to $Q$. This is related to the $Y$ system by
\be\label{Xseed}
Y_{i,s} = \prod_{i=1}^n X_{j,s}^{B_{ij}}.
\ee
In practice, instead of using the above relation, it is more convenient to exploit yet another relation with cluster algebra. Indeed, eqn.\eqref{Xseed} is the famous relation in between the $Y$ seed and the $X$ seed of the cluster algebra associated with $Q$. Let $\Xi\equiv\{k_1,\dots,k_s\}$ be the sequence of nodes associated with the finest fractional monodromy of the system. The frieze pattern associated to a 4D $\cn=2$ system is defined via $\Xi$ by applying eqn.\eqref{ratmap} to the $X$ seed. In other words, the rational functions $R^{(i)}_j$ get replaced by the corresponding $X$ seed mutations
\be\label{Xseedmut}
G^{(i)}_j(X_1,\dots,X_n)= \begin{cases} \frac{1}{X_i}\left(\prod_{k=1}^n X_k^{[B_{ik}]_+} + \prod_{k=1}^n X_k^{[B_{ki}]_+}\right)&\text{if } i = j\\ X_i &\text{else}\end{cases}
\ee
and the frieze pattern is obtained from the recursion
\be
X_{j,s+1} \equiv G_j(X_{1,s},\dots,X_{n,s})
\ee
where $G_j$ is the following rational map
\be
G_j(X_1,\dots,X_n) \equiv \mathscr{I}_{\sigma^{-1}} \circ G_j^{(k_{s})}\circ G_j^{(k_{s-1})}\circ \dots \circ G_j^{(k_{2})} \circ G_j^{(k_{1})}(X_1,\dots,X_n).
\ee
From eqn.\eqref{Xseedmut} we see that the $X$ seed frieze pattern is redundant whenever the finest $1/m$ fractional monodromy corresponds to an admissible sequence of nodes $\Xi$ that is supported on a strict subset of $Q_0$. Whenever this happens, typically, one can extract a finer frieze pattern out of the original one, we call this finer pattern the $Q$ system. The reason for this definition comes from the frieze pattern associated to pure $SU(n)$ SYM: the refined frieze pattern in this case is the well known $A_{n-1}$ type Kirillov--Reshetikin $Q$ system \cite{kirilov} (up to a twist of sign \cite{kedem}). Notice that, from eqn.\eqref{Xseedmut}, it follows that, whenever it exists, the $Q$ system depends only on $\text{card}(\Xi) < n$ variables.

\medskip

\noindent{\bf Remark.} In the math literature about cluster algebras, the frieze pattern is typically defined in terms of the mutation sequence that correspond, in our language, to the half monodromy. Let us denote it by $\widetilde{X}_{i,t}$. For the cluster algebras associated with 4D $\cn=2$ models, there is a finer frieze pattern, the $X_{i,s}$, corresponding to the minimal fractional monodromy. From eqn.\eqref{thebettersoln}, the two are related by $\widetilde{X}_{i,t} \equiv X_{i, t (k_{\Xi}+1) }$.

\medskip

From the discussion in the introduction follows that, if defined, the frieze pattern of an asymptotically free 4D $\cn=2$ system should satisfy linear recurrence relations. The form of these linear recurrences is going to depend on the specific charges of the various \textsc{susy} line defects involved. In particular, we expect to have different recursions associated to different generators of $\Gamma$, that are one to one with the nodes of the quiver.

\medskip

The presence of linear recurrence relations is indeed a leitmotif in the context of frieze patterns \cite{CoxeterConway,Coxeterf}. One of the biggest confirmations of the general theory that we have discussed above is given by the $SU(2)$ asymptotically free models. The recurrence relations, for this case, have been worked out in the math literature about frieze patterns. \cite{kellerrec} Another important example is given by the pure $SU(N)$ SYM theories. Indeed, for all the $Q$ systems of $A$ type there is a further reduction yielding precisely to a finite linear recurrence relation \cite{difrancesco}. In what follows, after reviewing the known results in the literature using the language that we have introduced so far, we perform a further test of our general ideas. As a byproduct of this fact, we construct infinitely many new families of $Q$ systems of $A$ type.

\subsection{The case of asymptotically free $SU(2)$ gauge theories}

Asymptotically free $SU(2)$ gauge theories are classified by the BPS quivers of affine $ADE$ type. 
The elements of low rank in the affine $ADE$ series correspond lagrangian asymptotically free models, increasing the ranks, 
we find $SU(2)$ SYM coupled to non canonical matter. More precisely the theories $SU(2)$ $N_f=0,1,2,3$ have BPS quivers equal respectively to $\widehat{A}(1,1)$, $\widehat{A}(2,1)$, $\widehat{A}(2,2)$, and $\widehat{D}_4$. Higher rank affine models correspond to $SU(2)$ SYM weakly gauging the diagonal flavor symmetry of a system of non--interacting $D_p$ Argyres--Douglas systems --- see table \ref{SU2AF}.

\begin{figure}
\begin{center}
\begin{equation}\label{Apqquivs}
A(p,q) \colon \quad\begin{gathered}
\xymatrix@R=0.7pc@C=0.7pc{
&q-1\ar[r]&q-2\ar[r]&\dots\ar[r]&1\ar[dr]&\\
q\ar[ur]\ar[dr]&&&&&0\\
&q+1\ar[r]&q+2\ar[r]&\dots\ar[r]&p+q-1\ar[ur]&\\}
\end{gathered}
\end{equation}

\bigskip

\begin{equation}
\widehat{D}_{r+2} \colon \quad\begin{gathered}
\xymatrix{
*++[o][F-]{r}\ar[dr]&&&&&& *++[o][F-]{1}\\
&r-1\ar[r]&r-2\ar[r]&\dots\ar[r]&3\ar[r]&2\ar[ur]\ar[dr]&\\
*++[o][F-]{r+1}\ar[ur]&&&&&&*++[o][F-]{0}\\}
\end{gathered} 
\end{equation}

\begin{equation}
\widehat{E}_6\colon \quad\begin{gathered}
\xymatrix{
&&*++[o][F-]{1}\ar[d]&&\\
&&2\ar[d]&&\\
*++[o][F-]{5} \ar[r]&6\ar[r]&7&4\ar[l]&*++[o][F-]{3}\ar[l]}
\end{gathered}
\end{equation}

\bigskip

\begin{equation}
\widehat{E}_7\colon \quad\begin{gathered}
\xymatrix{
&&&7\ar[d]&&\\
*++[o][F-]{1}\ar[r]&2\ar[r]&3\ar[r]&8&6\ar[l]&5\ar[l]&\ar[l]4}
\end{gathered}
\end{equation}

\bigskip

\begin{equation}
\widehat{E}_8\colon \quad\begin{gathered}
\xymatrix{
&&&&&6\ar[d]&&\\
*++[o][F-]{1}\ar[r]&2\ar[r]&3\ar[r]&4\ar[r]&5\ar[r]&9&8\ar[l]&\ar[l]7}
\end{gathered}
\end{equation}
\end{center}
\caption{Our conventions on the nodes of the affine quivers. For $\widehat{A}(p,q)$ each node is an extending one, for the other affine quivers above the extending nodes are circled.}\label{conventsaff}
\end{figure}

\begin{table}
\begin{center}
\begin{tabular}{|c|c|c|}
\hline
quiver & non canonical matter & $b$\\
\hline
$\widehat{A}(p,q)$&$D_p\oplus D_q$&\phantom{\bigg|}$2\,(\frac{1}{p} + \frac{1}{q})$\\
$\widehat{D}_{2+r}$&$D_2 \oplus D_2 \oplus D_{r}$&\phantom{\bigg|}$\frac{2}{r}$\\
$\widehat{E}_{3+r}$&$D_2 \oplus D_3 \oplus D_r$&\phantom{\bigg|}$\frac{6-r}{6r}$\\
\hline
\end{tabular}
\caption{$SU(2)$ asymptotically free theories. We denote by $D_2$ the free theory of two non--interacting $\cn=2$ hypers, while $D_3\simeq A_3$ Argyres--Douglas.}\label{SU2AF}
\end{center}
\end{table}

\medskip

Let us label the nodes of the affine quivers as in figure \ref{conventsaff}. Each of these model has a minimal BPS chamber corresponding to a stability condition that kills all but the simple objects in $\text{rep } \widehat{H}$. This stability condition has a half--monodromy given by the following sequence of nodes ($n\equiv\text{rank } \widehat{H}$):\footnote{ This weird choice for the labelings of the nodes is motivated by the simplicity of the recursion relations one gets.}
\bea
\Lambda_{\widehat{A}(p,q)} &\equiv \{ 0,1,\dots,q-1,p+q-1,p+q-2,\dots,q+2,q+1,q \}\\
\Lambda_{\widehat{D}_{r+2}} &\equiv \{0,1,2,\dots,r-1,r,r+1\}\qquad\qquad\Lambda_{\widehat{E}_{6}} \equiv \{7,6,4,2,5,3,1\}\\
\Lambda_{\widehat{E}_{7}} &\equiv \{8,7,6,3,5,4,1\}\qquad\qquad\qquad\qquad\Lambda_{\widehat{E}_8} \equiv \{9,8,6,5,7,4,3,2,1\}
\eea
The corresponding frieze pattern is elegantly written as follows:
\be
X_{i,s+1} \equiv \frac{1}{X_{i,s}} \left( 1 + \prod_{k \to i} X_{k,s} \prod_{i\to k} X_{k,s+1}  \right)
\ee
We have the following

\medskip

\noindent{\bf Theorem.} \cite{kellerrec} : \textit{If $Q$ is an affine quiver, then every frieze sequence $\{X_{i,s}\}_s$ satisfies a linear recurrence relation.}

\medskip

\textit{In particular, setting}
\be
\begin{cases} 
u_i \equiv i - q &0\leq i\leq q\\
u_i \equiv \text{max}(q-i,-q) &q<i\\
v_i \equiv - u_{i+q}
\end{cases}
\ee
\textit{the frieze sequence of type $\widehat{A}(p,q)$ satisfies}
\be
X_{i+q,s+u_i} - c \cdot X_{i,s} + X_{i-q,s+v_i} = 0
\ee
\textit{where the $i$ index is taken mod $p+q$. In particular, if $p=q$, as $i+q = i-q\text{ mod } 2q$, by iterating once the above, one obtains}
\be\label{affinerec3}
X_{i,s-q} - (c^2-2) X_{i,s} + X_{i,s+q} = 0.
\ee

\medskip

\textit{Moreover, \underline{let $i$ be an extending vertex} of $\widehat{H}$:}
\begin{itemize}
\item \textit{The $\widehat{D}_{r+2}$ frieze pattern satisfies:}
\be\label{affinerec1}
\begin{cases}
X_{i,s+r} - c \cdot X_{i,s} +X_{i,s-r}=0&r \text{ even}\\
X_{i,s+2r} - c \cdot  X_{i,s} +X_{i,s-2r}=0&r \text{ odd}\\
\end{cases}
\ee
\item \textit{The $\widehat{E}_{r+3}$ frieze pattern satisfies:}
\be\label{affinerec2}
X_{i,s-d_{r+3}} - c \cdot X_{i,s} + X_{i,s - d_{r+3}} = 0 \qquad (d_6,d_7,d_8)\equiv (6,12,30)
\ee
\end{itemize}

\noindent{\bf Remark.} It is intriguing to notice that in all of the eqns.\eqref{affinerec1},\eqref{affinerec2}, and \eqref{affinerec3}, the shift in the linear recursion is always given by the denominator $d_b$ of the coefficient $b$ of the beta function listed in table \ref{SU2AF}. In the appendix of \cite{kellerrec} one can find conjectural values for the order of the recursion relations corresponding to the non extending vertices for systems of type $\widehat{D}$ and $\widehat{E}$. While a relation with $d_b$ is manifest for the $\widehat{D}_{r+2}$ series (one finds only multiples of $d_b$), the same is not true for the three exceptional affine systems.

\subsection{The case of $SU(N+1)$ SYM}
All 4D $\cn=2$ pure SYM theories with simple lie group $G$ admit a BPS quiver description \cite{CNV,cattoy,nonsimply}. The IR Coulomb branch of the system is captured by the SW geometry
\be
e^{-Z} + e^{Z} + W_G(X,Y) + U^2 = 0 \subset C^4.
\ee
This model is a direct sum of the 2d $(2,2)$ $\widehat{A}(1,1)$ system with a minimal model of type $G$. The SYM models have an interesting square tensor product quiver representative in their mutation class. For $SU(N+1)$ SYM  such quiver is
\be\label{SYMAN}
\widehat{A}(1,1)\square A_N \colon
\begin{gathered}
\xymatrix{\circ_1\ar@<-0.4ex>[dd]\ar@<0.4ex>[dd]&&\circ_2\ar@<-0.4ex>[dd]\ar@<0.4ex>[dd]&&\circ_3\ar@<-0.4ex>[dd]\ar@<0.4ex>[dd]&&\circ_{N-1}\ar@<-0.4ex>[dd]\ar@<0.4ex>[dd]&&\circ_{N}\ar@<-0.4ex>[dd]\ar@<0.4ex>[dd]\\
&&&&&\cdots\\
\bullet_1\ar[uurr]&&\bullet_2\ar[uull]\ar[uurr]&&\bullet_3\ar[uull]&&\bullet_{N-1}\ar[uurr]&&\bullet_{N}\ar[uull]}\end{gathered}
\ee
This quiver admit factorized sequences of the form $A_N \coprod A_N$, and therefore has a fractional $1/2(N+1)$ monodromy: its sequence of nodes is
\be
\Xi\equiv \{ \bullet_1,\bullet_2,\dots,\bullet_{N-1},\bullet_N \}
\ee
while the corresponding permutation is
\be
\sigma \equiv \{(\bullet_1,\circ_1),(\bullet_2,\circ_2),\dots,(\bullet_{N-1},\circ_{N-1}),(\bullet_N,\circ_N)\} \in \mathfrak{S}_{2N}.
\ee
The full Coxeter factorized sequence the computes the half monodromy is
\be
\mathbf{m}_{\sigma^{N}(\Xi)} \circ \mathbf{m}_{\sigma^{N-1}(\Xi)} \circ \cdots \circ \mathbf{m}_{\sigma^2(\Xi)} \circ \mathbf{m}_{\sigma(\Xi)} \circ \mathbf{m}_{\Xi}.
\ee
and corresponds to a finte BPS chamber of the form $A_n \coprod A_N$. The frieze pattern associated with $(\Xi,\sigma)$ is
\be\label{SUNfrieze}
\begin{cases}
X_{\bullet_i,s+1} = X_{\circ_i,s}&i=1,\dots,N\\
X_{\circ_i,s+1} = (X^2_{\circ_i,s} + X_{\circ_{i-1},s}X_{\circ_{i+1},s})/X_{\bullet_i,s}&s\in\bZ\\
\text{with the convention: } X_{\circ_{0},s}\equiv X_{\circ_{N+1},s}\equiv 1.
\end{cases}
\ee
The $Q$ system is the sequence of $N$ variables 
\be
Q_{i,s} \equiv X_{\circ_i,s} \qquad i = 1,..,N \qquad s\in\mathbb{Z}
\ee
underlying the frieze pattern. The recursion relations inherited from eqn.\eqref{SUNfrieze} reads: 
\be\label{AnQsyss}
\begin{cases}Q_{i,s+1} Q_{i,s-1} = Q^2 _{i,s} + Q_{i-1,s} Q_{i+1,s}\\ Q_{0,s} \equiv Q_{N+1,s} \equiv 1\end{cases}
\ee
a sign--twisted version the Kirillov Reshtikin $Q$ system of type $A_N$ \cite{kedem}. Below we review the argument of \cite{difrancesco} that shows that the above $Q$ system can be solved entirely in terms of $Q_{1,s}$. Let us define $N$ matricies $i\times i$:
\be
(\cm_{i,s})_{ab} \equiv Q_{1,s+a+b-i-1} \qquad a,b = 1,\dots, i.
\ee
Let us denote $W_{i,s} \equiv |\cm_{i,s}| \equiv \text{det} (\cm_{i,s})$. From the Desnanot Jacobi formula about determinants and minors
\be
|M| \,|M^{1, n}_{1, n} |= |M^{n}_{1} |\,|M^{1}_{n}| - |M^{1}_{1}|\,|M^{n}_{n} |
\ee
applied to the $(i+1)\times (i+1)$ matrix
\be
(M)_{ab} \equiv Q_{1,s+a+b-i-2}\qquad a,b = 1,\dots,i+1
\ee
One obtains that
\be
W_{i,s+1} W_{i,s-1} = W_{i,s}^2 + W_{i+1,s}W_{i-1,s},
\ee
with the convention $W_{i,0} \equiv 1$. Since $W_{i,1} = Q_{i,1}$, this shows that
\be\label{WQAn}
W_{i,s}\equiv Q_{i,s}.
\ee
In other words, also the $Q$ system is redundant: it can be solved entirely in terms of only one of its variables, namely $Q_{1,s}$. Let us define
\be
F_s \equiv Q_{1,s} \qquad s\in\mathbb{Z}
\ee
The identification in eqn.\eqref{WQAn} entails that $W_{i,s}$ satisfy the other boundary condition for the $Q$ system: in other words, the $(N+1)\times (N+1)$ matrix $\cm_{N+1,s}$ satisfies
\be\label{bundry}
W_{N+1,s} = 1 \qquad \forall \, s\in \bZ
\ee
In particular,
\bea
0 &= W_{N+1,s} - W_{N+1,s+1}\\
& = 
\left|\begin{matrix} F_{s - N} & F_{s-N+1}&\cdots&F_s\\ 
F_{s - N+1} & F_{s-N+2}&\cdots&F_{s+1}\\
\vdots&\vdots& &\vdots\\
F_{s}&F_{s+1}&\cdots&F_{s+N}\end{matrix}\right| - \left|\begin{matrix} F_{s - N+1} & F_{s-N+2}&\cdots&F_{s+1}\\ 
F_{s - N+2} & F_{s-N+3}&\cdots&F_{s+2}\\
\vdots&\vdots&&\vdots\\
F_{s+1}&F_{s+2}&\cdots&F_{s+N+1}\end{matrix}\right|
\eea
and since the two matrices above share $N$ identical rows, the two determinants can be combined by multilinearity. The resulting matrix has vanishing determinant: its columns have to be linearly dependent. Each row contributes essentially the same relation, namely
\be\label{SUNrex}
F_{s} + (-1)^{N+1} F_{s + N + 1} + c_1 \, F_{s+1} +  c_2 \, F_{s+2} + \cdots +  c_{N-1} \, F_{s+N-1} + c_{N} \, F_{s+N} = 0.
\ee
In conclusion in this way we obtain a linear recurrence relation for the sequence $F_s$ with $N+2$ terms.

\subsection{The case of $\widehat{A}(p,p)\,\boxtimes \,A_N$}
\begin{figure}
$$
\begin{gathered}
\begin{xy} 0;<0.6pt,0pt>:<0pt,-0.6pt>:: 
(45,260) *+{\circ_{1,1}} ="0",
(0,170) *+{\bullet_{1,1}} ="1",
(0,80) *+{\circ_{1,2}} ="2",
(45,0) *+{\bullet_{1,2}} ="3",
(100,110) *+{\circ_{1,3}} ="4",
(100,210) *+{\bullet_{1,3}} ="5",
(195,260) *+{\bullet_{2,1}} ="6",
(150,170) *+{\circ_{2,1}} ="7",
(240,210) *+{\circ_{2,3}} ="8",
(150,80) *+{\bullet_{2,2}} ="9",
(240,110) *+{\bullet_{2,3}} ="10",
(195,0) *+{\circ_{2,2}} ="11",
(345,260) *+{\circ_{3,1}} ="12",
(300,170) *+{\bullet_{3,1}} ="13",
(400,210) *+{\bullet_{3,3}} ="14",
(300,80) *+{\circ_{3,2}} ="15",
(400,110) *+{\circ_{3,3}} ="16",
(345,0) *+{\bullet_{3,2}} ="17",
"0", {\ar"1"},
"0", {\ar"5"},
"6", {\ar"0"},
"2", {\ar"1"},
"1", {\ar"7"},
"2", {\ar"3"},
"9", {\ar"2"},
"4", {\ar"3"},
"3", {\ar"11"},
"4", {\ar"5"},
"10", {\ar"4"},
"5", {\ar"8"},
"7", {\ar"6"},
"8", {\ar"6"},
"6", {\ar"12"},
"7", {\ar"9"},
"13", {\ar"7"},
"8", {\ar"10"},
"14", {\ar"8"},
"11", {\ar"9"},
"9", {\ar"15"},
"11", {\ar"10"},
"10", {\ar"16"},
"17", {\ar"11"},
"12", {\ar"13"},
"12", {\ar"14"},
"15", {\ar"13"},
"16", {\ar"14"},
"15", {\ar"17"},
"16", {\ar"17"},
\end{xy}
\end{gathered}
$$
\caption{The quiver of the system $A(3,3)\boxtimes A_3$.}\label{333ex}
\end{figure}
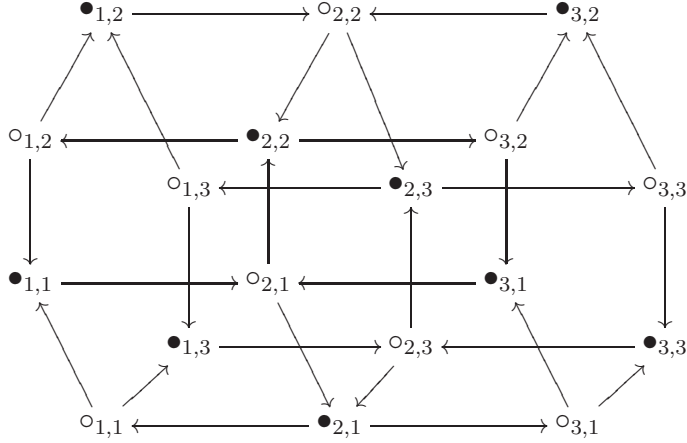

Paralleling the analogy in between the $SU(2)$ example we have discussed in the introduction and the affine models, that represent $SU(2)$ coupled in a UV asymptotically free fashion with some non canonical matter system, one expect the same type of behavior by coupling $G$ SYM to some non canonical `nice' matter system, where `nice' means that the central charge $k_G$ associated to the two point function of the $G$ flavor current is such that $k_G \leq 2 h(G)$. A large variety of models of this kind were introduced in reference\cite{infinitelymany}. In between these systems, the ones that have a combinatorics that more closely resembles the one of $G$ SYM is obtained by Type II B engineering on the following geometry
\be
 e^{-pZ} +  e^{pZ} + W_G(X,Y) + U^2 = 0.
\ee
The BPS quiver of these system is $\widehat{A}(p,p)\,\square \,A_N$. We draw in figure \ref{333ex} an example.
It is clear for example that the $1/m$ fractional monodromy of order $m=\text{lcm}(2,h)$ of the system is obtained from the mutation sequence on the $\bullet_{i,a}$ nodes with permutation $\sigma\equiv \{(\bullet_i,\circ_i)\}$. Notice that each $X_{\bullet_{i,a}}$ has only 4 neighbors, but for the boundary elements $X_{\bullet_{i,1}}$ and $X_{\bullet_{i,N}}$ that have 3. Mutation of the $X$ seed is a local operation, therefore the effect of $\mathbf{m}_\Xi$ is easily decoded. We obtain the following frieze pattern:
\be
\begin{cases}
X_{\bullet_{i,a},s+1} = X_{\circ_{i,a},s}&a=1,...,N\\
X_{\circ_{i,a},s+1}=\frac{1}{X_{\bullet_{i,a},s}}(X_{\circ_{i-1,a},s}X_{\circ_{i+1,a},s} +X_{\circ_{i,a-1},s}X_{\circ_{i,a+1},s} )&i=1,...,p\\
X_{\circ_{i,0},s} \equiv X_{\circ_{i,N+1},s}\equiv 1&s\in\bZ\\
X_{\circ_{i+p,a},s}\equiv X_{\circ_{i,a},s}
\end{cases}
\ee
The corresponding $Q$ system is
\be
Q_{i,a,s} \equiv X_{\circ_{i,a,s}}
\ee
and we obtain
\be\label{QAPPA}
\begin{cases}
Q_{i,a,s+1}Q_{i,a,s-1}=Q_{i-1,a,s}Q_{i+1,a,s} +Q_{i,a-1,s}Q_{i,a+1,s} &a=1,...,N\\
Q_{i,0,s} \equiv Q_{i,N+1,s}\equiv 1&i=1,...,p\\
Q_{i+p,a,s}\equiv Q_{i,a,s}&s\in\bZ\\
\end{cases}
\ee
Notice the compatibility with the $N=1$ case: Indeed, in this case we obtain precisely the $Q$ system of type $\widehat{A}(p,p)$. Also in this case the $Q$ system can be solved in terms of a subset of its variables as in the previous section. Let us define
\be
F^i_{s} \equiv Q_{i,1,s}\quad\text{thus}\quad F^{i+p}_s = F^{i}_s
\ee
Then, construct the following $a\times a$ matricies:
\be
(\cm_{i,a,s})_{mn} \equiv F^{i-m+n}_{s+m+n-a-1} \qquad m,n=1,...,a\quad i=1,...,p
\ee
again by the Desnanot--Jacobi formula applied to the $(a+1)\times (a+1)$ matrix
\be
(M^{(i)})_{mn} \equiv F^{i-m+n}_{s+m+n-a-2}\qquad m,n = 1,\dots,a+1
\ee
 we see that the $W_{i,a,s}\equiv \text{det}(\cm_{i,a,s})$ obey the $Q$ system. Set $W_{i,0,s} \equiv1$. Notice that $W_{i,1,s} = F^i_s$. Therefore the $W_{i,a,s}$ obey the $Q$ system of eqn.\eqref{QAPPA}. In particular, it follows that $W_{i,N+1,s}=1$ for all $i$'s:
\be
W_{i,N+1,s}=\left|\begin{matrix} F^i_{s - N} & F^{i+1}_{s-N+1}&\cdots&F^{i+N+1}_s\\ 
F^{i-1}_{s - N+1} & F^i_{s-N+2}&\cdots&F^{i+N}_{s+1}\\
\vdots&\vdots& &\vdots\\
F^{i-N}_{s}&F^{i-N+1}_{s+1}&\cdots&F^{i}_{s+N}\end{matrix}\right| = 1
\ee
By the same mechanism as above, the whole family 
\be
\{F^1_s,F^2_s,\cdots,F^p_s\}
\ee
obeys a linear recursion relation of a form similar to eqn.\eqref{SUNrex}, obtained using the identities $i=1,\dots,p$
\be\label{differew}
W_{i-1,N+1,s} - W_{i,N+1,s+1} = 0.
\ee
Exploiting the fact that the corresponding $(N+1)\times (N+1)$ matrices share $N$ columns we infer the linear relation, namely
\be
F^{i}_{s+N+1} + (-1)^{N+1} \, F^{i-N-1}_s = \sum_{a=1}^N c_a \, F^{i-N+a}_{s+a}.
\ee

\section*{Ackowledgements}
We thank the organizers of the Simons Center Summer 2013 workshop ``Defects'' at the Simons Center in Stony Brook, where this project has been conceived. The work of MDZ is supported by the NSF grant PHY-1067976.

\appendix

\section{Arnol'd $Y$ systems}\label{arn}

\begin{table}[hbp!]
\begin{center}
\begin{tabular}{|c|c|c|c|c|}\hline
name & polynomial $W(x,y,z)$ & weights $q_i$&$\ell$ & Coxeter--Dynkin diagram \\\hline
%
$\begin{matrix}\\ E_{13}\\ \phantom{a}\end{matrix}$ & $\begin{matrix}\\ x^3+xy^5+z^2\\ \phantom{a}\end{matrix}$ & $\begin{matrix}\\ 1/3, 2/15, 1/2\\ \phantom{a}\end{matrix}$ &7&{\scriptsize
$\begin{gathered}\xymatrix{\bullet \ar@{-}[r]\ar@{-}[d]\ar@{..}[dr]&\bullet \ar@{-}[r]\ar@{-}[d]\ar@{..}[dr]&\bullet \ar@{-}[r]\ar@{-}[d]\ar@{..}[dr]&\bullet \ar@{-}[r]\ar@{-}[d]\ar@{..}[dr]&\bullet \ar@{-}[r]\ar@{-}[d]\ar@{..}[dr]&\bullet \ar@{-}[d]\ar@{..}[dr]&\\
\bullet \ar@{-}[r]&\bullet \ar@{-}[r]&\bullet \ar@{-}[r]&\bullet \ar@{-}[r]&\bullet \ar@{-}[r]&\bullet\ar@{-}[r]&\bullet }\end{gathered}$}\\\hline
%
%
$\begin{matrix}\\ Z_{11}\\ \phantom{a}\end{matrix}$ & $\begin{matrix}\\ x^3y+y^5+z^2\\ \phantom{a}\end{matrix}$ & $\begin{matrix}\\ 4/15, 1/5, 1/2\\ \phantom{a}\end{matrix}$ &7&{\scriptsize
$\begin{gathered}\xymatrix{\bullet \ar@{-}[r]\ar@{-}[d]&\bullet \ar@{-}[r]\ar@{-}[d]\ar@{..}[dl]&\bullet \ar@{-}[d]\ar@{..}[dl] & \\
\bullet \ar@{-}[r]\ar@{-}[d]\ar@{..}[dr]&\bullet \ar@{-}[r]\ar@{-}[d]\ar@{..}[dr]&\bullet \ar@{-}[r]\ar@{-}[d]\ar@{..}[dr]&\bullet \ar@{-}[d]\\
\bullet \ar@{-}[r]&\bullet \ar@{-}[r]&\bullet \ar@{-}[r]&\bullet\\
}\end{gathered}$}\\\hline
$\begin{matrix}\\ Z_{12}\\ \phantom{a}\end{matrix}$ & $\begin{matrix}\\ x^3y+xy^4+z^2\\ \phantom{a}\end{matrix}$ & $\begin{matrix}\\ 3/11, 2/11, 1/2\\ \phantom{a}\end{matrix}$ & 5 &{\scriptsize
$\begin{gathered}\xymatrix{\bullet \ar@{-}[r]\ar@{-}[d]&\bullet \ar@{-}[r]\ar@{-}[d]\ar@{..}[dl]&\bullet \ar@{-}[d]\ar@{..}[dl] & &\\
\bullet \ar@{-}[r]\ar@{-}[d]\ar@{..}[dr]&\bullet \ar@{-}[r]\ar@{-}[d]\ar@{..}[dr]&\bullet \ar@{-}[r]\ar@{-}[d]\ar@{..}[dr]&\bullet \ar@{-}[d] \ar@{..}[dr]&\\
\bullet \ar@{-}[r]&\bullet \ar@{-}[r]&\bullet \ar@{-}[r]&\bullet \ar@{-}[r]&\bullet\\
}\end{gathered}$}\\\hline
$\begin{matrix}\\ Z_{13}\\ \phantom{a}\end{matrix}$ & $\begin{matrix}\\ x^3y+y^6+z^2\\ \phantom{a}\end{matrix}$ & $\begin{matrix}\\ 5/18, 1/6, 1/2\\ \phantom{a}\end{matrix}$ & 8 &{\scriptsize
$\begin{gathered}\xymatrix{\bullet \ar@{-}[r]\ar@{-}[d]&\bullet \ar@{-}[r]\ar@{-}[d]\ar@{..}[dl]&\bullet \ar@{-}[d]\ar@{..}[dl] & &\\
\bullet \ar@{-}[r]\ar@{-}[d]\ar@{..}[dr]&\bullet \ar@{-}[r]\ar@{-}[d]\ar@{..}[dr]&\bullet \ar@{-}[r]\ar@{-}[d]\ar@{..}[dr]&\bullet \ar@{-}[d] \ar@{..}[dr]\ar@{-}[r]&\bullet \ar@{-}[d]\\
\bullet \ar@{-}[r]&\bullet \ar@{-}[r]&\bullet \ar@{-}[r]&\bullet \ar@{-}[r]&\bullet\\
}\end{gathered}$}\\\hline
%

%
$\begin{matrix}\\ W_{13}\\ \phantom{a}\end{matrix}$ & $\begin{matrix}\\ x^4+xy^4+z^2\\ \phantom{a}\end{matrix}$ & $\begin{matrix}\\ 1/4, 3/16, 1/2\\ \phantom{a}\end{matrix}$ & 7 &
${\scriptsize\begin{gathered}\xymatrix{\bullet \ar@{-}[r]\ar@{-}[d]&\bullet \ar@{-}[r]\ar@{-}[d]\ar@{..}[dl]&\bullet \ar@{-}[d]\ar@{..}[dl]\ar@{-}[r]& \bullet \ar@{-}[d]\ar@{..}[dl] &\\
\bullet \ar@{-}[r]\ar@{-}[d]\ar@{..}[dr]&\bullet \ar@{-}[r]\ar@{-}[d]\ar@{..}[dr]&\bullet \ar@{-}[r]\ar@{-}[d]\ar@{..}[dr]&\bullet \ar@{-}[d] \ar@{..}[dr]&\\
\bullet \ar@{-}[r]&\bullet \ar@{-}[r]&\bullet \ar@{-}[r]&\bullet \ar@{-}[r]&\bullet\\
}\end{gathered}}$\\\hline
%
%
$\begin{matrix}\\ Q_{11}\\ \phantom{a}\end{matrix}$ & $\begin{matrix}\\ x^2z+y^3+yz^3\\ \phantom{a}\end{matrix}$ & $\begin{matrix}\\ 7/18, 1/3, 2/9\\ \phantom{a}\end{matrix}$ & 8 &{\scriptsize
$\begin{gathered}\xymatrix{\bullet\ar@{-}[r]\ar@{-}@/_/[dd]&\bullet \ar@{..}[ddl]\ar@{-}@/_/[dd]&&\\
\bullet \ar@{-}[r]\ar@{-}[d]&\bullet \ar@{-}[d]\ar@{..}[dl]& &\\
\bullet \ar@{-}[r]\ar@{-}[d]\ar@{..}[dr]&\bullet \ar@{-}[r]\ar@{-}[d]\ar@{..}[dr]&\bullet \ar@{-}[d] \ar@{..}[dr]& \\
\bullet \ar@{-}[r]&\bullet \ar@{-}[r]&\bullet \ar@{-}[r]&\bullet\\
}\end{gathered}$}\\\hline
%
%
\end{tabular}
\caption{Arnol'd's $14$ exceptional singularities that are not of the form $W_G+W_{G^\prime}$}\label{arnoldtable}
\end{center}
\label{arnoldtable2}
\end{table}%

\begin{table*}
\begin{center}
\begin{tabular}{|c|c|c|c|c|}\hline
name & polynomial $W(x,y,z)$ & weights $q_i$ &$\ell$& Coxeter--Dynkin diagram \\\hline
$\begin{matrix}\\ S_{11}\\ \phantom{a}\end{matrix}$ & $\begin{matrix}\\ x^2z+yz^2+y^4\\ \phantom{a}\end{matrix}$ & $\begin{matrix}\\ 5/16, 1/4, 3/8\\ \phantom{a}\end{matrix}$ &7&
${\scriptsize\begin{gathered}\xymatrix{\bullet\ar@{-}[r]\ar@{-}@/_/[dd]&\bullet \ar@{..}[ddl]\ar@{-}@/_/[dd]&\\
\bullet \ar@{-}[r]\ar@{-}[d]&\bullet \ar@{-}[d]\ar@{..}[dl]\ar@{-}[r]&\bullet \ar@{-}[d]\ar@{..}[dl]\\
\bullet \ar@{-}[r]\ar@{-}[d]\ar@{..}[dr]&\bullet \ar@{-}[r]\ar@{-}[d]\ar@{..}[dr]&\bullet \ar@{-}[d] \\
\bullet \ar@{-}[r]&\bullet \ar@{-}[r]&\bullet\\
}\end{gathered}}$\\\hline
$\begin{matrix}\\ S_{12}\\ \phantom{a}\end{matrix}$ & $\begin{matrix}\\ x y^3 + x^2 z + y z^2\\ \phantom{a}\end{matrix}$ & $\begin{matrix}\\ 4/13, 3/13, 5/13\\ \phantom{a}\end{matrix}$ &11&{\scriptsize
$\begin{gathered}\xymatrix{\bullet\ar@{-}[r]\ar@{-}@/_/[dd]&\bullet \ar@{..}[ddl]\ar@{-}@/_/[dd]&\\
\bullet \ar@{-}[r]\ar@{-}[d]&\bullet \ar@{-}[d]\ar@{..}[dl]\ar@{-}[r]&\bullet \ar@{-}[d]\ar@{..}[dl]\\
\bullet \ar@{-}[r]\ar@{-}[d]\ar@{..}[dr]&\bullet \ar@{-}[r]\ar@{-}[d]\ar@{..}[dr]&\bullet \ar@{-}[d]\ar@{..}[dr] \\
\bullet \ar@{-}[r]&\bullet \ar@{-}[r]&\bullet \ar@{-}[r]&\bullet\\
}\end{gathered}$}\\\hline

\end{tabular}
\end{center}
\label{arnoldtable3}
\caption{Arnol'd's $14$ exceptional singularities that are not of the form $W_G+W_{G^\prime}$ --- continued.}
\end{table*}%

Arnol'd 4D $\cn=2$ SCFT's are superconformal systems obtained by Type II B superstring engineering on singular Calabi Yau hypersurfaces of $\C^4$ of the form $W(X,Y,Z,V) \equiv 0$ where $W$ is a quasi homogenous singularity such that the corresponding LG model has central charge $\hat{c}<2$, as required by 2d/4D correspondence. In this appendix we draw from \cite{Arnold1,Arnold2} the tables about quasi homogenous unimodal and bimodal singularities that meet the criterion of \cite{GVW}. The BPS quiver for all of these models can be obtained by 2d/4D correspondence from the Coxeter--Dynkin diagram associated to the singularity \cite{Arnold1}: the latter encodes the stokes matrix $S$ of the corresponding 2d $tt^*$ structure. In particular we list all our predictions about the order of the quantum monodromy operator $\ell$. All these models admit Coxeter factorized sequences of mutations that can be used to extract the corresponding BPS spectrum and periodic $Y$ systems \cite{Arnold1,Arnold2}.

\begin{table}
\begin{center}
\begin{tabular}{|c|c|c|}\hline
name & $W(x,y,z)$ & Coxeter--Dynkin diagram\\
\hline
$\begin{matrix}\\ E_{19}\\ \phantom{a}\end{matrix}$ & $\begin{matrix}\\ x^3+xy^7+z^2\\ \phantom{a}\end{matrix}$ & {\scriptsize $\begin{gathered}\xymatrix{
\bullet \ar@{-}[r]\ar@{-}[d]\ar@{..}[dr]&\bullet \ar@{-}[r]\ar@{-}[d]\ar@{..}[dr]&\bullet \ar@{-}[r]\ar@{-}[d]\ar@{..}[dr]&\bullet \ar@{-}[r]\ar@{-}[d]\ar@{..}[dr]&\bullet \ar@{-}[r]\ar@{-}[d]\ar@{..}[dr]&\bullet \ar@{-}[d]\ar@{..}[dr]\ar@{-}[r]&\bullet \ar@{-}[d]\ar@{..}[dr]\ar@{-}[r]&\bullet \ar@{-}[d]\ar@{..}[dr]\ar@{-}[r]&\bullet\ar@{-}[d]\ar@{..}[dr]\\
\bullet \ar@{-}[r]&\bullet \ar@{-}[r]&\bullet \ar@{-}[r]&\bullet \ar@{-}[r]&\bullet \ar@{-}[r]&\bullet\ar@{-}[r]&\bullet\ar@{-}[r]&\bullet\ar@{-}[r]&\bullet\ar@{-}[r]&\bullet }\end{gathered}$}\\\hline
$\begin{matrix}\\ Z_{17}\\ \phantom{a}\end{matrix}$ & $\begin{matrix}\\ x^3y+y^8+z^2\\ \phantom{a}\end{matrix}$ & {\scriptsize $\begin{gathered}\xymatrix{
\bullet \ar@{-}[r]\ar@{-}[d]&\bullet \ar@{-}[r]\ar@{-}[d]\ar@{..}[dl]&\bullet \ar@{-}[d]\ar@{..}[dl] & &\\
\bullet \ar@{-}[r]\ar@{-}[d]\ar@{..}[dr]&\bullet \ar@{-}[r]\ar@{-}[d]\ar@{..}[dr]&\bullet \ar@{-}[r]\ar@{-}[d]\ar@{..}[dr]&\bullet \ar@{-}[r]\ar@{-}[d]\ar@{..}[dr]&\bullet \ar@{-}[r]\ar@{-}[d]\ar@{..}[dr]&\bullet \ar@{-}[d] \ar@{..}[dr]\ar@{-}[r]&\bullet \ar@{-}[d]\\
\bullet \ar@{-}[r]&\bullet \ar@{-}[r]&\bullet \ar@{-}[r]&\bullet \ar@{-}[r]&\bullet \ar@{-}[r]&\bullet \ar@{-}[r]&\bullet\\
}\end{gathered}$}\\\hline
$\begin{matrix}\\ Z_{18}\\ \phantom{a}\end{matrix}$ & $\begin{matrix}\\ x^3y+xy^6+z^2\\ \phantom{a}\end{matrix}$ & {\scriptsize $\begin{gathered}\xymatrix{
\bullet \ar@{-}[r]\ar@{-}[d]&\bullet \ar@{-}[r]\ar@{-}[d]\ar@{..}[dl]&\bullet \ar@{-}[d]\ar@{..}[dl] & &\\
\bullet \ar@{-}[r]\ar@{-}[d]\ar@{..}[dr]&\bullet \ar@{-}[r]\ar@{-}[d]\ar@{..}[dr]&\bullet \ar@{-}[r]\ar@{-}[d]\ar@{..}[dr]&\bullet \ar@{-}[r]\ar@{-}[d]\ar@{..}[dr]&\bullet \ar@{-}[r]\ar@{-}[d]\ar@{..}[dr]&\bullet \ar@{-}[d] \ar@{..}[dr]\ar@{-}[r]&\bullet \ar@{-}[d]\ar@{..}[dr]\\
\bullet \ar@{-}[r]&\bullet \ar@{-}[r]&\bullet \ar@{-}[r]&\bullet \ar@{-}[r]&\bullet \ar@{-}[r]&\bullet \ar@{-}[r]&\bullet\ar@{-}[r]&\bullet\\
}\end{gathered}$}\\\hline
$\begin{matrix}\\ Z_{19}\\ \phantom{a}\end{matrix}$ & $\begin{matrix}\\ x^3y+y^9+z^2\\ \phantom{a}\end{matrix}$ &  {\scriptsize $\begin{gathered}\xymatrix{
\bullet \ar@{-}[r]\ar@{-}[d]&\bullet \ar@{-}[r]\ar@{-}[d]\ar@{..}[dl]&\bullet \ar@{-}[d]\ar@{..}[dl] & &\\
\bullet \ar@{-}[r]\ar@{-}[d]\ar@{..}[dr]&\bullet \ar@{-}[r]\ar@{-}[d]\ar@{..}[dr]&\bullet \ar@{-}[r]\ar@{-}[d]\ar@{..}[dr]&\bullet \ar@{-}[r]\ar@{-}[d]\ar@{..}[dr]&\bullet \ar@{-}[r]\ar@{-}[d]\ar@{..}[dr]&\bullet \ar@{-}[d] \ar@{..}[dr]\ar@{-}[r]&\bullet \ar@{-}[d] \ar@{..}[dr]\ar@{-}[r]&\bullet \ar@{-}[d]\\
\bullet \ar@{-}[r]&\bullet \ar@{-}[r]&\bullet \ar@{-}[r]&\bullet \ar@{-}[r]&\bullet \ar@{-}[r]&\bullet \ar@{-}[r]&\bullet \ar@{-}[r]&\bullet\\
}\end{gathered}$}\\\hline
$\begin{matrix}\\ W_{17}\\ \phantom{a}\end{matrix}$ & $\begin{matrix}\\ x^4+x y^5+z^2\\ \phantom{a}\end{matrix}$ &  {\scriptsize $\begin{gathered}\xymatrix{\bullet \ar@{-}[r]\ar@{-}[d]&\bullet \ar@{-}[r]\ar@{-}[d]\ar@{..}[dl]&\bullet \ar@{-}[d]\ar@{..}[dl]\ar@{-}[r]&\bullet \ar@{-}[d]\ar@{..}[dl]\ar@{-}[r]& \bullet \ar@{-}[d]\ar@{..}[dl]\\
\bullet \ar@{-}[r]\ar@{-}[d]\ar@{..}[dr]&\bullet \ar@{-}[r]\ar@{-}[d]\ar@{..}[dr]&\bullet \ar@{-}[r]\ar@{-}[d]\ar@{..}[dr]&\bullet \ar@{-}[r]\ar@{-}[d]\ar@{..}[dr]&\bullet \ar@{-}[r]\ar@{-}[d]\ar@{..}[dr]&\bullet \ar@{-}[d]\\
\bullet \ar@{-}[r]&\bullet \ar@{-}[r]&\bullet \ar@{-}[r]&\bullet \ar@{-}[r]&\bullet \ar@{-}[r]&\bullet\\
}\end{gathered}$}\\\hline
\end{tabular}
\caption{Exceptional bimodal singularities that are not of the form $W_G + W_{G^\prime}$.}\label{ArnB1}
\end{center}
\end{table}

\begin{table}
\begin{center}
\begin{tabular}{|c|c|c|}\hline
%
$\begin{matrix}\\ Q_{17}\\ \phantom{a}\end{matrix}$ & $\begin{matrix}\\ x^3+yz^2+xy^5\\ \phantom{a}\end{matrix}$ &{\scriptsize$\begin{gathered}\xymatrix{\bullet\ar@{-}[r]\ar@{-}@/_/[dd]&\bullet \ar@{..}[ddl]\ar@{-}@/_/[dd]&&\\
\bullet \ar@{-}[r]\ar@{-}[d]&\bullet \ar@{-}[d]\ar@{..}[dl]& &\\
\bullet \ar@{-}[r]\ar@{-}[d]\ar@{..}[dr]&\bullet \ar@{-}[r]\ar@{-}[d]\ar@{..}[dr]&\bullet \ar@{-}[r]\ar@{-}[d]\ar@{..}[dr]&\bullet \ar@{-}[r]\ar@{-}[d]\ar@{..}[dr]&\bullet \ar@{-}[r]\ar@{-}[d]\ar@{..}[dr]&\bullet\ar@{-}[d]\ar@{..}[dr] \\
\bullet \ar@{-}[r]&\bullet \ar@{-}[r]&\bullet \ar@{-}[r]&\bullet \ar@{-}[r]&\bullet \ar@{-}[r]&\bullet \ar@{-}[r]&\bullet\\
}\end{gathered}$}\\\hline
%
%
$\begin{matrix}\\ S_{16}\\ \phantom{a}\end{matrix}$ & $\begin{matrix}\\ x^2z+yz^2+xy^4\\ \phantom{a}\end{matrix}$ & {\scriptsize$\begin{gathered}\xymatrix{\bullet\ar@{-}[r]\ar@{-}@/_/[dd]&\bullet \ar@{..}[ddl]\ar@{-}@/_/[dd]&\\
\bullet \ar@{-}[r]\ar@{-}[d]&\bullet \ar@{-}[d]\ar@{..}[dl]\ar@{-}[r]&\bullet \ar@{-}[d]\ar@{..}[dl]\ar@{-}[r]&\bullet \ar@{-}[d]\ar@{..}[dl]&\\
\bullet \ar@{-}[r]\ar@{-}[d]\ar@{..}[dr]&\bullet \ar@{-}[r]\ar@{-}[d]\ar@{..}[dr]&\bullet \ar@{-}[r]\ar@{-}[d]\ar@{..}[dr]&\bullet \ar@{-}[r]\ar@{-}[d]\ar@{..}[dr]&\bullet \ar@{-}[d] \\
\bullet \ar@{-}[r]&\bullet \ar@{-}[r]&\bullet \ar@{-}[r]&\bullet \ar@{-}[r]&\bullet\\
}\end{gathered}$}\\\hline
$\begin{matrix}\\ S_{17}\\ \phantom{a}\end{matrix}$ & $\begin{matrix}\\ x^2z+yz^2+y^6\\ \phantom{a}\end{matrix}$ & {\scriptsize$\begin{gathered}\xymatrix{\bullet\ar@{-}[r]\ar@{-}@/_/[dd]&\bullet \ar@{..}[ddl]\ar@{-}@/_/[dd]&\\
\bullet \ar@{-}[r]\ar@{-}[d]&\bullet \ar@{-}[d]\ar@{..}[dl]\ar@{-}[r]&\bullet \ar@{-}[d]\ar@{..}[dl]\ar@{-}[r]&\bullet \ar@{-}[d]\ar@{..}[dl]\ar@{-}[r]&\bullet \ar@{-}[d]\ar@{..}[dl]\\
\bullet \ar@{-}[r]\ar@{-}[d]\ar@{..}[dr]&\bullet \ar@{-}[r]\ar@{-}[d]\ar@{..}[dr]&\bullet \ar@{-}[r]\ar@{-}[d]\ar@{..}[dr]&\bullet \ar@{-}[r]\ar@{-}[d]\ar@{..}[dr]&\bullet \ar@{-}[d] \\
\bullet \ar@{-}[r]&\bullet \ar@{-}[r]&\bullet \ar@{-}[r]&\bullet \ar@{-}[r]&\bullet\\
}\end{gathered}$}\\\hline
\end{tabular}
\caption{Exceptional bimodal singularities that are not of the form $W_G + W_{G^\prime}$ --- continued.}
\end{center}
\end{table}

\begin{table}
\begin{center}
\begin{tabular}{|c|c|c|}\hline
name & $W(x,y,z)$ & Coxeter--Dynkin diagram\\
\hline
$\begin{matrix}\\ Z_{1,0}  \\ \phantom{a}\end{matrix}$ {\footnotesize (15)} & $\begin{matrix}\\ x^3 y+y^7+z^2 \\ \phantom{a}\end{matrix}$ & {\scriptsize $\begin{gathered}\xymatrix{
\bullet \ar@{-}[r]\ar@{-}[d]&\bullet \ar@{-}[r]\ar@{-}[d]\ar@{..}[dl] &\bullet \ar@{-}[d]\ar@{..}[dl]&\\
\bullet \ar@{-}[r]\ar@{-}[d]\ar@{..}[dr]&\bullet \ar@{-}[r]\ar@{-}[d]\ar@{..}[dr]&\bullet \ar@{-}[r]\ar@{-}[d]\ar@{..}[dr]&\bullet \ar@{-}[r]\ar@{-}[d]\ar@{..}[dr]&\bullet \ar@{-}[r]\ar@{-}[d]\ar@{..}[dr]&\bullet \ar@{-}[d]\\
\bullet \ar@{-}[r]&\bullet \ar@{-}[r]&\bullet \ar@{-}[r]&\bullet \ar@{-}[r]&\bullet \ar@{-}[r]&\bullet\\
}\end{gathered}$}\\\hline
%
%
$\begin{matrix}\\ S_{1,0} \\ \phantom{a}\end{matrix}$ {\footnotesize (14)}& $\begin{matrix}\\ x^2 z + y z^2 + y^5 \\ \phantom{a}\end{matrix}$ &{\scriptsize$\begin{gathered}\xymatrix{\bullet\ar@{-}[r]\ar@{-}@/_/[dd]&\bullet \ar@{..}[ddl]\ar@{-}@/_/[dd]&\\
\bullet \ar@{-}[r]\ar@{-}[d]&\bullet \ar@{-}[d]\ar@{..}[dl]\ar@{-}[r]&\bullet \ar@{-}[d]\ar@{..}[dl]\ar@{-}[r]&\bullet \ar@{-}[d]\ar@{..}[dl]&\\
\bullet \ar@{-}[r]\ar@{-}[d]\ar@{..}[dr]&\bullet \ar@{-}[r]\ar@{-}[d]\ar@{..}[dr]&\bullet \ar@{-}[r]\ar@{-}[d]\ar@{..}[dr]&\bullet \ar@{-}[d] \\
\bullet \ar@{-}[r]&\bullet \ar@{-}[r]&\bullet \ar@{-}[r]&\bullet\\
}\end{gathered}$}\\\hline
$\begin{matrix}\\ U_{1,0}  \\ \phantom{a}\end{matrix}$ {\footnotesize (14)} & $\begin{matrix}\\ x^3 + x z^2 + x y^3 \\ \phantom{a}\end{matrix}$ & {\scriptsize$\begin{gathered}\xymatrix{
\bullet\ar@{-}[r]\ar@{-}@/_/[dd]&\bullet \ar@{..}[ddl]\ar@{-}@/_/[dd]\ar@{-}[r]&\bullet \ar@{..}[ddl]\ar@{-}@/_/[dd]&\\
\bullet \ar@{-}[r]\ar@{-}[d]&\bullet \ar@{-}[d]\ar@{..}[dl]\ar@{-}[r]&\bullet \ar@{-}[d]\ar@{..}[dl]& \\
\bullet \ar@{-}[r]\ar@{-}[d]\ar@{..}[dr]&\bullet \ar@{-}[r]\ar@{-}[d]\ar@{..}[dr]&\bullet \ar@{-}[r]\ar@{-}[d]\ar@{..}[dr]&\bullet\ar@{-}[d] \\
\bullet \ar@{-}[r]&\bullet \ar@{-}[r]&\bullet \ar@{-}[r]&\bullet\\
}\end{gathered}$}\\\hline
\end{tabular}
\caption{Quasi--homogeneous elements of the 8 infinite series of bimodal singularities that are not of the form $W_G + W_{G^\prime}$. We indicate the corresponding Milnor numbers in parenthesis.}
\end{center}
\end{table}

\begin{table}
\begin{center}
\begin{tabular}{|c|c|c|c|}\hline
 & $q_x, q_y, q_z$ & $\hat{c}$ & $\ell$ \\\hline
$E_{19}$ & 1/3 , 1/7 , 1/2 & 8/7 & 18 \\
$Z_{17}$ & 7/24, 1/8, 1/2 & 7/6 & 10 \\
$Z_{18}$ & 5/17, 2/17, 1/2 & 20/17 & 14 \\
$Z_{19}$ & 8/27, 1/9, 1/2 & 32/27 & 22 \\
$W_{17}$ & 1/4, 3/20, 1/2 & 6/5 & 8 \\
$Q_{17}$ & 1/3, 2/15, 13/30 & 6/5 & 12 \\
$S_{16}$ & 5/17, 3/17, 7/17 & 21/17 & 13 \\
$S_{17}$ & 7/24, 1/6, 5/12 & 5/4 & 9 \\\hline
$Z_{1,0}$ & 2/7, 1/7, 1/2 & 8/7 & 6 \\
$S_{1,0}$& 3/10, 1/5, 2/5 & 6/5 & 4 \\
$U_{1,0}$& 1/3, 2/9, 1/3 & 11/9 & 9 \\\hline
\end{tabular}
\end{center}
\caption{Chiral charges $q_i$, central charges $\hat{c}$ and orders $\ell$ of 
the quantum monodromy $\mathscr{M}_q$ for the quasi--homogeneous bimodal singularities listed in the tables above.}\label{numerology}
\end{table}

\newpage

\newpage

\newpage

\end{document}